\documentclass[sigconf]{acmart}
\AtBeginDocument{%
  \providecommand\BibTeX{{%
    \normalfont B\kern-0.5em{\scshape i\kern-0.25em b}\kern-0.8em\TeX}}}

\copyrightyear{2025}
\acmYear{2025}
\makeatletter
\def\@ACM@copyright@check@cc{}
\makeatother
\setcopyright{cc}
\setcctype{by}
\acmConference[CHI '25]{CHI Conference on Human Factors in Computing Systems}{April 26-May 1, 2025}{Yokohama, Japan}
\acmBooktitle{CHI Conference on Human Factors in Computing Systems (CHI '25), April 26-May 1, 2025, Yokohama, Japan}\acmDOI{10.1145/3706598.3713392}
\acmISBN{979-8-4007-1394-1/25/04}

\usepackage{float}
\usepackage{subcaption}
\usepackage{multirow}
\usepackage{makecell}

\begin{document}

\title{Modeling the Impact of Visual Stimuli on Redirection Noticeability with Gaze Behavior in Virtual Reality}

\author{Zhipeng Li}
\affiliation{%
  \institution{Tsinghua Univeristy}
  \city{Beijing}
  \country{China}
}
\affiliation{%
  \institution{ETH Zürich}
  \city{Zürich}
  \country{Switzerland}
}
\email{zhipeng.li@inf.ethz.ch}

\author{Yishu Ji}
\affiliation{%
  \institution{Georgia Institute of Technology}
  \city{Atlanta}
  \state{Georgia}
  \country{USA}
}
\email{yji329@gatech.edu}

\author{Ruijia Chen}
\affiliation{%
  \institution{University of Wisconsin-Madison}
  \city{Madison}
  \state{Wisconsin}
  \country{USA}
}
\email{ruijia.chen@wisc.edu}

\author{Tianqi Liu}
\affiliation{%
  \institution{Cornell University}
  \city{Ithaca}
  \state{New York}
  \country{USA}
}
\email{tl889@cornell.edu}

\author{Yuntao Wang}
\authornote{Corresponding author}
\affiliation{%
  \institution{Key Laboratory of Pervasive Computing, Ministry of Education, Tsinghua University}
  \city{Beijing}
  \country{China}
}
\email{yuntaowang@tsinghua.edu.cn}

\author{Yuanchun Shi}
\affiliation{%
  \institution{Tsinghua University}
  \city{Beijing}
  \country{China}
}
\email{shiyc@tsinghua.edu.cn}

\author{Yukang Yan}
\affiliation{%
  \institution{University of Rochester}
  \city{Rochester}
  \state{New York}
  \country{USA}
}
\email{yanyukanglwy@gmail.com}

\renewcommand{\shortauthors}{Li, et al.}
\begin{abstract}

While users could embody virtual avatars that mirror their physical movements in Virtual Reality, these avatars' motions can be redirected to enable novel interactions.
Excessive redirection, however, could break the user's sense of embodiment due to perceptual conflicts between vision and proprioception. 
While prior work focused on avatar-related factors influencing the noticeability of redirection, we investigate how the visual stimuli in the surrounding virtual environment affect user behavior and, in turn, the noticeability of redirection.
Given the wide variety of different types of visual stimuli and their tendency to elicit varying individual reactions, 
we propose to use users' gaze behavior as an indicator of their response to the stimuli and model the noticeability of redirection.
We conducted two user studies to collect users' gaze behavior and noticeability, investigating the relationship between them and identifying the most effective gaze behavior features for predicting noticeability. 
Based on the data, we developed a regression model that takes users' gaze behavior as input and outputs the noticeability of redirection. 
We then conducted an evaluation study to test our model on unseen visual stimuli, achieving an accuracy of 0.012 MSE. 
We further implemented an adaptive redirection technique and conducted a preliminary study to evaluate its effectiveness with complex visual stimuli in two applications. 
The results indicated that participants experienced less physical demanding and a stronger sense of body ownership when using our adaptive technique, demonstrating the potential of our model to support real-world use cases.
\end{abstract}



\begin{teaserfigure}
  \includegraphics[width=\textwidth]{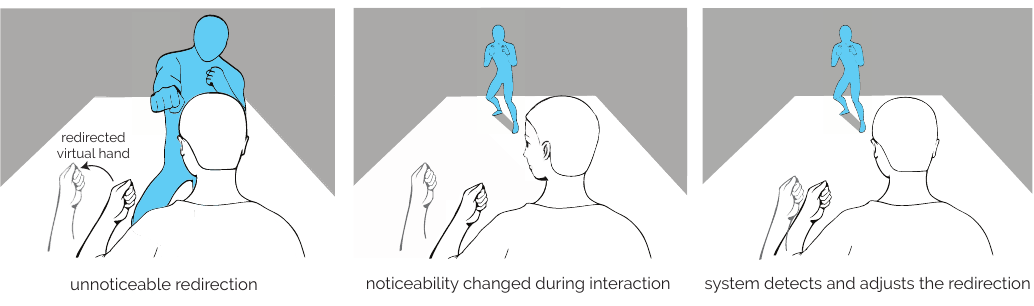}
  \caption{
    In this paper, we explored the impact of visual stimuli on the noticeability of redirection in Virtual Reality. 
    We developed a computational model that takes users' gaze behavior as input and predicts the noticeability of redirection under different visual stimuli.
    Using this model, we implemented an adaptive redirection technique, demonstrated in a boxing training scenario:
    Left: When the opponent approaches and attacks, visual stimuli are intense, making the redirection unnoticable.
    Middle: As the opponent retreats, visual stimuli decrease that causes the noticeability becoming higher during the interaction.
    Right: When the model detects the change in noticeability, the system dynamically adjusts the redirection magnitude, ensuring it remains unnoticed.
  }
  \label{fig:teaser}
\end{teaserfigure}


\maketitle
\section{Introduction}
\label{section:introduction}

Virtual Reality (VR) systems enable users to embody virtual avatars by mirroring their physical movements and aligning their perspective with virtual avatars' in real time. 
As the head-mounted displays (HMDs) block direct visual access to the physical world, users primarily rely on visual feedback from the virtual environment and integrate it with proprioceptive cues to control the avatar’s movements and interact within the VR space.
Since human perception is heavily influenced by visual input~\cite{gibson1933adaptation}, 
VR systems have the unique capability to control users' perception of the virtual environment and avatars by manipulating the visual information presented to them.
Leveraging this, various redirection techniques have been proposed to enable novel VR interactions, 
such as redirecting users' walking paths~\cite{razzaque2005redirected, suma2012impossible, steinicke2009estimation},
modifying reaching movements~\cite{gonzalez2022model, azmandian2016haptic, cheng2017sparse, feick2021visuo},
and conveying haptic information through visual feedback to create pseudo-haptic effects~\cite{samad2019pseudo, dominjon2005influence, lecuyer2009simulating}.
Such redirection techniques enable these interactions by manipulating the alignment between users' physical movements and their virtual avatar's actions.


However, these redirection techniques are most effective when the manipulation remains undetected~\cite{gonzalez2017model, li2022modeling}. 
If the redirection becomes too large, the user may not mitigate the conflict between the visual sensory input (redirected virtual movement) and their proprioception (actual physical movement), potentially leading to a loss of embodiment with the virtual avatar and making it difficult for the user to accurately control virtual movements to complete interaction tasks~\cite{li2022modeling, wentzel2020improving, feuchtner2018ownershift}. 
While proprioception is not absolute, users only have a general sense of their physical movements and the likelihood that they notice the redirection is probabilistic. 
This probability of detecting the redirection is referred to as \textbf{noticeability}~\cite{li2022modeling, zenner2024beyond, zenner2023detectability} and is typically estimated based on the frequency with which users detect the manipulation across multiple trials.


Prior research has explored factors influencing the noticeability of redirected motion, including the redirection's magnitude~\cite{wentzel2020improving, poupyrev1996go}, direction~\cite{li2022modeling, feuchtner2018ownershift}, and the visual characteristics of the virtual avatar~\cite{ogawa2020effect, feick2024impact}.
While these factors focus on the avatars, the surrounding virtual environment can also influence the users' behavior and in turn affect the noticeability of redirection.
This, however, remains underexplored.
One such prominent external influence is through the visual channel - the users' visual attention is constantly distracted by complex visual effects and events in practical VR scenarios.
We thus want to investigate how \textbf{visual stimuli in the virtual environment} affect the noticeability of redirection.
With this, we hope to complement existing works that focus on avatars by incorporating environmental visual influences to enable more accurate control over the noticeability of redirected motions in practical VR scenarios.

Since each visual event is a complex choreography of many underlying factors (type of visual effect, location, duration, etc.), it is extremely difficult to quantify or parameterize visual stimuli.
Furthermore, individuals respond differently to even the same visual events.
Prior neuroscience studies revealed that factors like age, gender, and personality can influence how quickly someone reacts to visual events~\cite{gillon2024responses, gale1997human}. 
Therefore, aiming to model visual stimuli in a way that is generalizable and applicable to different stimuli and users, we propose to use users' \textbf{gaze behavior} as an indicator of how they respond to visual stimuli.
In this paper, we used various gaze behaviors, including gaze location, saccades~\cite{krejtz2018eye}, fixations~\cite{perkhofer2019using}, and the Index of Pupil Activity (IPA)~\cite{duchowski2018index}.
These behaviors indicate both where users are looking and their cognitive activity, as looking at something does not necessarily mean they are attending to it.
Our goal is to investigate how these gaze behaviors stimulated by various visual stimuli relate to the noticeability of redirection.
With this, we contribute a model that allows designers and content creators to adjust the redirection in real-time responding to dynamic visual events in VR.

To achieve this, we conducted user studies to collect users' noticeability of redirection under various visual stimuli.
To simulate realistic VR scenarios, we adopted a dual-task design in which the participants performed redirected movements while monitoring the visual stimuli.
Specifically, participants' primary task was to report if they noticed an offset between the avatar's movement and their own, while their secondary task was to monitor and report the visual stimuli.
As realistic virtual environments often contain complex visual effects, we started with simple and controlled visual stimulus to manage the influencing factors.

We first conducted a confirmation study (N=16) to test whether applying visual stimuli (opacity-based) actually affects their noticeability of redirection. 
The results showed that participants were significantly less likely to detect the redirection when visual stimuli was presented $(F_{(1,15)}=5.90,~p=0.03)$.
Furthermore, by analyzing the collected gaze data, results revealed a correlation between the proposed gaze behaviors and the noticeability results $(r=-0.43)$, confirming that the gaze behaviors could be leveraged to compute the noticeability.

We then conducted a data collection study to obtain more accurate noticeability results through repeated measurements to better model the relationship between visual stimuli-triggered gaze behaviors and noticeability of redirection.
With the collected data, we analyzed various numerical features from the gaze behaviors to identify the most effective ones. 
We tested combinations of these features to determine the most effective one for predicting noticeability under visual stimuli.
Using the selected features, our regression model achieved a mean squared error (MSE) of 0.011 through leave-one-user-out cross-validation. 
Furthermore, we developed both a binary and a three-class classification model to categorize noticeability, which achieved an accuracy of 91.74\% and 85.62\%, respectively.

To evaluate the generalizability of the regression model, we conducted an evaluation study (N=24) to test whether the model could accurately predict noticeability with new visual stimuli (color- and scale-based animations).
Specifically, we evaluated whether the model's predictions aligned with participants' responses under these unseen stimuli.
The results showed that our model accurately estimated the noticeability, achieving mean squared errors (MSE) of 0.014 and 0.012 for the color- and scale-based visual stimili, respectively, compared to participants' responses.
Since the tested visual stimuli data were not included in the training, the results suggested that the extracted gaze behavior features capture a generalizable pattern and can effectively indicate the corresponding impact on the noticeability of redirection.

Based on our model, we implemented an adaptive redirection technique and demonstrated it through two applications: adaptive VR action game and opportunistic rendering.
We conducted a proof-of-concept user study (N=8) to compare our adaptive redirection technique with a static redirection, evaluating the usability and benefits of our adaptive redirection technique.
The results indicated that participants experienced less physical demand and stronger sense of embodiment and agency when using the adaptive redirection technique. 
These results demonstrated the effectiveness and usability of our model.

In summary, we make the following contributions.
\begin{itemize}
    \item 
    We propose to use users' gaze behavior as a medium to quantify how visual stimuli influences the noticebility of redirection. 
    Through two user studies, we confirm that visual stimuli significantly influences noticeability and identify key gaze behavior features that are closely related to this impact.
    \item 
    We build a regression model that takes the user's gaze behavioral data as input, then computes the noticeability of redirection.
    Through an evaluation study, we verify that our model can estimate the noticeability with new participants under unseen visual stimuli.
    These findings suggest that the extracted gaze behavior features effectively capture the influence of visual stimuli on noticeability and can generalize across different users and visual stimuli.
    \item 
    We develop an adaptive redirection technique based on our regression model and implement two applications with it.
    With a proof-of-concept study, we demonstrate the effectiveness and potential usability of our regression model on real-world use cases.

\end{itemize}

\section{Related work}
\label{section:related_work}

\subsection{Redirection in VR}

As users tend to prioritize visual information over other sensory channels when they are facing various information from the sensory system, (i.e., visual dominance~\cite{rock1964vision, gibson1933adaptation}), VR provides the opportunity to manipulate the visual information that users perceive to enable novel interactions.
While the manipulation is applied to users' movement and remains undetected, users will fall into the illusion that makes them believe their physical body movement is consistent with the manipulated virtual movement, which is called redirection.
Redirection can be implemented by adding an offset to the user's movement in VR to adjust the trajectory slightly~\cite{kohli2012redirected, gonzalez2023sensorimotor}.

Redirection has been widely used in VR applications to improve interaction performance and enable new interactions, including visuo-haptic illusion, augmenting input techniques and redirected walking.
As one of the most frequently-used forms of body input, hand movement has been widely explored as the subject of redirection illusions.
Hand redirection has been employed to alter the perceived shape \cite{redirectedtouching, zhao2018functional} and location \cite{HapticRetargeting, cheng2017sparse} of passive haptic props; this creates visuo-haptic illusions~\cite{yu2020pseudo}, which have been found to increase users' reutilization of physical counterparts to different virtual objects. 
To improve the interaction efficiency, researchers applied redirection techniques by adding offsets to users' hand movement~\cite{frees2007prism, montano2017erg}.
As one of the earliest works that modified the user's body movement to enhance input performance, the Go-Go technique~\cite{poupyrev1996go} extended the virtual hand's depth with a non-linear function to enable users to interact with objects beyond the reach.
Ownershift~\cite{feuchtner2018ownershift} proposed a technique that allowed the user’s physical hand to shift to comfortable poses while keeping the virtual hand in mid-air, to keep the user unaware of the movement and avoid physical fatigue.
\citeauthor{wentzel2020improving} proposed a hand position amplification technique with an adaptive function, enabling users to interact with objects beyond reach while keeping the offset unnoticed~\cite{wentzel2020improving}.
Prior work explored manipulating the movement of other body parts, e.g.,
redirected walking techniques to guide users to specific physical locations while walking in the virtual environment.
By applying slightly angular offsets to users' footsteps, this technique allows users to walk in a boundless virtual environment within the confines of a restricted physical space~\cite{RDWroom-scale, rietzler2020telewalk}.

These redirection-based interaction techniques enable novel functionalities in VR (e.g., redirected walking) or improve users' interaction performance (e.g., Go-Go technique).
These studies highlight the application and benefit of redirection in VR interactions, which motivates us to further explore redirection techniques in VR.

\subsection{Noticeability of redirection in VR}
Though redirection-based methods enable various novel interactions in VR, previous studies suggested that it is also important to main embodiment during redirection~\cite{wentzel2020improving, zenner2023detectability, zenner2019estimating}.
The challenges around maintaining embodiment with redirection techniques are in how to minimize users' noticeability of offsets between their own bodies and virtual avatars.
To investigate the detection threshold of redirection, \citeauthor{burns2006hand} implemented redirection motion techniques in a game scenario and derived a detection threshold of 19.1 degrees (19cm) between the real and virtual hand~\cite{burns2006hand}.
Similarly, \citeauthor{lee2015enlarging} investigated the threshold for finger tracking errors and derived a much lower Just-Noticeable Difference (JND) of 5.2 cm, using a dot to indicate the fingertip position~\cite{lee2015enlarging} rather than a full representation of virtual hands.
To extend the detection threshold to a noticing probability, \citeauthor{li2022modeling} studied noticeability of redirection with different strength and direction on the user's arm movement and provided a model to compute the noticeability for given offset strength and direction~\cite{li2022modeling}.

In addition, adding coherent haptic feedback to the user's motion can also impact the noticeability of applied motion offset.
\citeauthor{abtahi2018visuo} investigated the fingertip offset detection threshold along with a physical proxy providing haptic feedback~\cite{abtahi2018visuo}.
The derived thresholds achieved 49.5 degrees on the horizontal axis, which was larger than the previously reported value.
Similary, \citeauthor{feick2021visuo} investigated how to leverage simple physical proxies to provide visuo-haptic illusions and investigated the noticeability of discrepancy between the physical and virtual object.
Their results indicated that users could bear a bigger offset when they gained more sensory information from other modalities.

Another effective approach to making redirection less noticeable is to manipulate the virtual environment by leveraging users' moments of inattention or blindness.
One method involves performing these manipulations outside the user’s field of view. 
For example, \citeauthor{suma2010exploiting} altered the geometry of a virtual room behind the user to subtly redirect walking paths~\cite{suma2010exploiting}. 
Similarly, \citeauthor{lohse2019leveraging} and \citeauthor{patras2022body} remapped virtual objects to physical props for haptic retargeting when they were outside the user’s view~\cite{lohse2019leveraging, patras2022body}.
Another approach is to introduce manipulations within the user’s field of view but outside their focus of attention. 
\citeauthor{marwecki2019mise} developed a system that uses eye tracking and attention models to apply changes only when objects fall outside the user’s visual attention~\cite{marwecki2019mise}.
However, these manipulations primarily focused on altering the virtual environment, rather than redirecting the movement of virtual avatars. 
In the context of virtual motion redirection, \citeauthor{zenner2023detectability} proposed applying virtual hand position offsets during user blinks. 
Their findings revealed that detection thresholds were significantly higher when the saccade direction opposed the hand offset direction~\cite{zenner2023detectability, zenner2021blink}.
While \citeauthor{zenner2023detectability} proposed redirecting users' motions during blinks, we explored the extent to which this redirection can be applied and examined its noticeability.

These studies reveal that the noticeability of redirection in VR can be influenced by various factors. 
While much of the research has focused on virtual avatars, the impact of the surrounding virtual environment on noticeability remains largely unexplored. 
To address this gap, we propose to investigate and model how visual stimuli affect the noticeability of redirection in this paper.

\subsection{Gaze behaviors for HCI}

Gaze behaviors have become crucial for understanding users' mental states and interaction intentions, especially with the integration of eye tracking in HMDs and smart glasses.
Beyond indicating where users are looking, gaze behaviors have also been used to classify attentional directions and indicate users' cognitive states~\cite{wang2019exploring}.
For instance, \citeauthor{benedek2017eye} demonstrated that pupil dilation is linked to cognitive focus~\cite{benedek2017eye}, while \citeauthor{duchowski2018index} introduced the Index of Pupil Activity (IPA) as a metric for cognitive load, which has been applied in HCI applications such as adaptive MR user interfaces~\cite{lindlbauer2019context}.
Furthermore, \citeauthor{annerer2021reliably} emphasized the role of pupil dilation in differentiating between internal and external attention~\cite{annerer2021reliably}.
In addition to pupil features, saccadic eye movements (saccade) and fixation duration are key indicators of cognitive load.
\citeauthor{zagermann2016measuring} and \citeauthor{holmqvist2011eye} found that longer fixations and shorter saccades were linked to higher cognitive demands~\cite{zagermann2016measuring, holmqvist2011eye}. 
These findings suggest a strong correlation between gaze behaviors and cognitive activity. 

Previous studies suggest that users' cognitive activities influence gaze behaviors over several seconds, rather than just a few frames. 
For example, \citeauthor{faber2018automated} recently demonstrated that content-independent gaze features with a 12-second window were effective for estimating cognitive load during reading tasks~\cite{faber2018automated}. 
Similarly, a time window-based method was proved to be effective in tasks such as film watching~\cite{mills2016automatic}, interactive tutoring~\cite{hutt2016eyes}, and lecture viewing~\cite{hutt2017gaze}. 
These studies also indicate that shorter windows (less than 10 seconds) may not capture enough fixation and saccade information to accurately detect covert inattention~\cite{bixler2016automatic, hutt2016eyes} which lead to lower accuracy~\cite{hutt2017gaze, hutt2016eyes}. 
Therefore, a longer windows (20-30 seconds) were more suitable for using gaze behaviors for estimating users' cognitive activities.

Based on the findings of previous studies, we propose to investigate how to use the gaze behaviors (gaze saccade, fixation, pupil activity) to compute the noticeability of redirection under various visual stimuli.

\section{Methodology}
\label{section:methodology}

To explore how to use gaze behaviors to compute the noticeability of redirection under visual stimuli, we employed a dual-task design in the following confirmation and data collection studies, which allows us to collet noticeability responses from participants while simultaneously presenting visual stimuli. 
In this section, we detail the methodology step by step.

\subsection{Selecting left arm for investigation}

Previous studies have applied redirection techniques to users' walking paths~\cite{kohli2012redirected}, hand positions~\cite{wentzel2020improving}, and arm motions~\cite{poupyrev1996go}.
Considering that the arms are among the most frequently used body parts, we focused our investigation on the redirection of arm movements.
To control for the influence of hand dominance, all user studies were conducted on left arms of right-handed participants.
We acknowledge that hand dominance and different body parts may lead to noticeability difference of the applied redirection, we believe that this approach and the main findings will be both applicable to the right arm and extendable to other body parts in future studies.

\subsection{Redirection mechanism}

In this paper, we adopted the same redirection mechanism as previous studies~\cite{li2022modeling}, which applied angular redirection to users' elbow joints during movement. 
The redirection was applied dynamically, starting with no redirection at the initial pose (pointing to the ground) and gradually increasing to the maximum redirection at the target ending pose.
The redirection of the intermediate motion was calculated based on the relative angular distance from the starting pose and was adjusted linearly throughout the movement.
The strength of the redirection was adjusted by modifying the maximum redirection applied at the ending pose.

\subsection{Dual-task design}

In realistic virtual environments, users often engage with complex visual effects (e.g., game props) while controlling their virtual avatars. 
To simulate this, we designed a dual-task study where participants were required to perform redirection motions (primary task) while monitoring and responding to visual stimuli at the same time (secondary task).

\subsubsection{Primary task.}
As the primary task, participants were asked to perform arm motions with redirected virtual arms in VR. 
Each motion was defined by a starting arm pose and a target ending pose. 
The starting pose was fixed in a natural resting position, with the arm positioned beside the body. 
The target ending pose was sampled from a motion capture dataset that included daily life poses, such as walking, sports, sitting, and others, as described in ~\autoref{section:target_poses}.
To ensure that all participants performed the movements consistently, we added an intermediate checkpoint pose between the starting and ending poses. 
During the study, both the target ending pose and the intermediate checkpoint pose were rendered as semi-transparent, allowing participants to observe them without causing visual occlusion. 
In contrast, the participants' redirected virtual arm was rendered normally, as illustrated in~\autoref{figure:formalapparatus}.

At the beginning of each trial, participants were instructed to perform the starting pose, in which they lowered their physical arm and pointed to the floor. 
Since there was no redirection at this position, the virtual arm also pointed downward. 
For each trial, a semi-transparent intermediate pose and target pose were displayed. 
Participants were then instructed to lift their physical arms, guiding their virtual arms past the intermediate pose to reach the target pose. 
Throughout this process, participants were asked to keep the virtual arm within their field of view at all times.
After reaching the target pose, participants were asked to move their virtual arm back to the starting pose and report to experimenters verbally whether they perceived any difference between their physical and virtual movement, according to the yes/no paradigm~\cite{leek2001adaptive}.
We estimated the noticeability of redirection in each condition with the ratio of positive responses (indicating noticed redirection in the trials) to the number of trials, referring to previous studies~\cite{li2022modeling}.

\subsubsection{Secondary task.}
In parallel with the primary task, participants were asked to monitor visual stimuli that appeared within their field of view. 
The stimuli consisted of a simple animation on a virtual sphere, presented alongside the virtual avatar. 
For example, in the opacity-based stimuli condition, the animation began with a fully transparent sphere, gradually increased to full opacity, and then returned to transparency.
The location and duration of the animation were adjusted to control the intensity of the visual stimuli between trials, following previous studies suggesting that these properties influence the intensity~\cite{li2024predicting}. 
The virtual sphere moved in sync with the participant's head movements, maintaining the same relative position within their field of view.
To prevent participants from predicting the timing of the stimuli, the animation began at a random moment after the trial started and repeated at random intervals (ranging from 1 to 3 seconds). 
Participants were instructed to press a button on a controller held by their right hand as soon as they noticed the animation was starting.

\subsection{Sampling target poses}
\label{section:target_poses}
For the ending poses, we selected 25 distinct poses from the CMU MoCap dataset~\cite{CMUMocap}. 
To ensure the diversity of poses, these poses were selected based on clustered subsets using the HDBSCAN algorithm~\cite{leland2017hdbscan}, based on the skeletal distance function proposed in~\cite{shakhnarovich05learning}, calculated as:

\begin{equation}
    \text{Distance}(\alpha_1, \alpha_2) = \max \limits_{1 \leq i \leq L} \sum_{d \in x, y, z} (\alpha^{i}_{d, 1} - \alpha^{i}_{d, 2})
\end{equation}

where $L$ represents the number of joints in the pose, and $x, y, z$ denote the spatial coordinates of each joint. 
This function determines the maximum skeletal distance between two poses $\alpha_{1}$ and $\alpha_{2}$ across all joints.
The sampled poses are displayed in \autoref{appendix:poses}.

\section{Confirmation Study}
\label{section:formativestudy}

Although it seems evident that adding additional visual stimuli may distract users and influence the noticeability of redirection, we conducted a confirmation study to validate this hypothesis and assess the effectiveness of our dual-task design.

\subsection{Design}



We employed a factorial study design to manage both independent and control variables.


\subsubsection{Independent variables}
In this study, we aimed to investigate whether applying visual stimuli affects noticeability. 
Therefore, our initial independent variable was the presence or absence of visual stimuli. 
To further explore the impact of various visual stimuli, we extended the independent variable to the intensity of visual stimuli, ranging from none to high.
We manipulated intensity by adjusting the duration and placement of virtual animations, following previous studies~\cite{gutwin2017peripheral, li2024predicting}. 
Through a pilot study, we identified three levels of duration: Short (0.2 sec), Medium (1 sec), and Long (2 sec).
For placement, we defined three layout configurations: Sparse (stimuli appear only in the corner areas), Median (stimuli appear in both the corner and peripheral areas), and Dense (stimuli appear throughout the entire field of view), as shown in~\autoref{figure:formalapparatus}.
In each layout, we randomly picked one candidate to animate the visual stimuli.
Additionally, we included a baseline condition with no visual stimuli.
The order of these conditions was randomized.


\subsubsection{Control variables}
We varied the magnitude and direction of the redirection as control variables. 
Based on the results from related research~\cite{li2022modeling}, we set the redirection magnitude from 0 to 30 degrees with an interval of 5 degrees, which covers the from being unnoticeable (no redirection) to easily noticeable.
We also varied the direction of redirection, sampling both horizontal and vertical directions.
As a result, each participant completed $(3~durations~\times~3~layouts~+~1~baseline)$  $\times (7$ redirection magnitudes $\times 2$ redirection direction - 1) $= 130$ trials in total.
The order of all redirection magnitudes and directions was randomized.


\subsubsection{Dependent variables}
The noticeability of redirection was recorded as the primary dependent variable in this study and was estimated with the proportion of positive responses across all trials for each condition where redirection was applied.
Additionally, we captured participants' gaze behavior data with the HMD's eye tracker.


\begin{figure}[t]
    \centering
    \includegraphics[width=0.9\columnwidth]{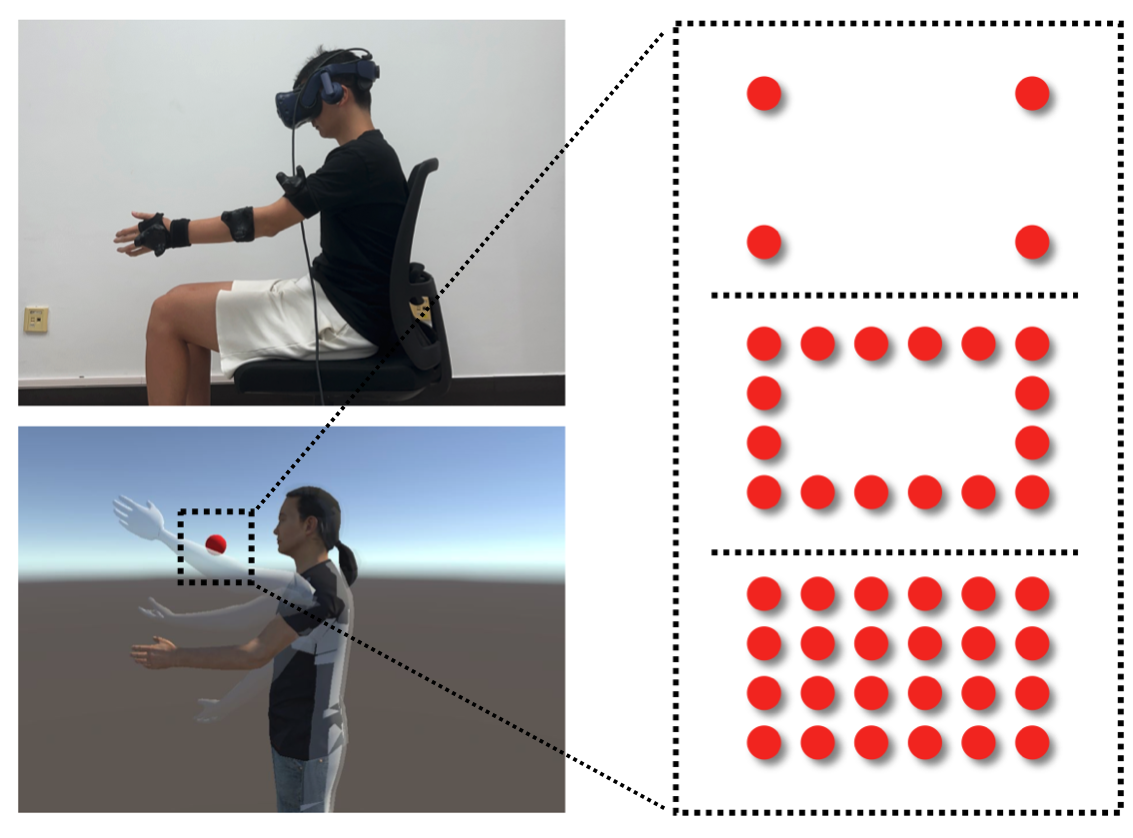}
    \caption{The apparatus of the formative study. 
    Wearing a headset, the participant wears three motion trackers to track their arm pose and sit on a comfortable chair. 
    While the virtual avatar mirrors the arm movement of the participant, the participant observes the virtual avatar's movement from a first-person point of view and follows the semi-transparent checkpoint pose to reach the semi-transparent target pose.
    As the secondary task, a virtual animation will start with different durations and locations.
    The right figure illustrates the possible locations of the red ball, named Sparse, Median, and Dense, accordingly.}
    \label{figure:formalapparatus}
\end{figure}

\subsection{Participants \& Apparatus}
The participants (N = 16) were recruited through an online questionnaire from a local university. 
The participants (7 females, 9 males) had an average age of 21.25 years ($SD = 1.71$). 
All were trichromats and right-handed. 
Prior to the experiment, participants self-evaluated on their familiarity with VR, reporting an average score of $3.75\ (SD=0.75) $on a 7-point Likert scale (1 - not at all familiar, 4 - neutral, 7 - very familiar).

We implemented the experimental application in VR with a HTC Vive pro headset in Unity 2019, powered by an Intel Core i7 CPU and an NVIDIA GeForce RTX 3080 GPU. 
Throughout the experimental sessions, participants were seated and equipped with three Vive Trackers affixed to their left shoulder, elbow, and waist using nylon straps.
Based on data given by the tracker, we reconstructed the left arm movement on a virtual humanoid avatar from the Microsoft RocketBox avatar library~\cite{gonzalez2020rocketbox} with the user’s viewpoint coinciding with the avatar’s (as shown in \autoref{figure:formalapparatus}).
All gaze data was recorded with the HTC Viveo pro built-in gaze tracker.
All statistical analyses were conducted with SPSS 26.0.


\subsection{Procedure}

To avoid bias from the participants knowing that we were intentionally introducing redirection, we introduced the purpose of the study as an evaluation of a motion capture and reconstruction technique and clarified the real purpose to participants after the study.
Participants were first provided with a walk-through of the platform.
Then, participants were provided with a warm-up session to ensure that they were familiar with the primary and secondary tasks.
After that, each participant completed 10 sessions ($3~durations~\times~3~layouts~+~1~baseline$) of experiments.
They took 2-minute breaks after every two sessions to reduce fatigue.
We recorded the participant's behavioral data, including the position and orientation of hand, elbow, shoulder, gaze, and pupil dilation, at a rate of 60 Hz.
The study lasted around 40 minutes and each participant was compensated with 15 US dollars.


\begin{figure}[t]
    \centering
    \includegraphics[width=0.9\columnwidth]{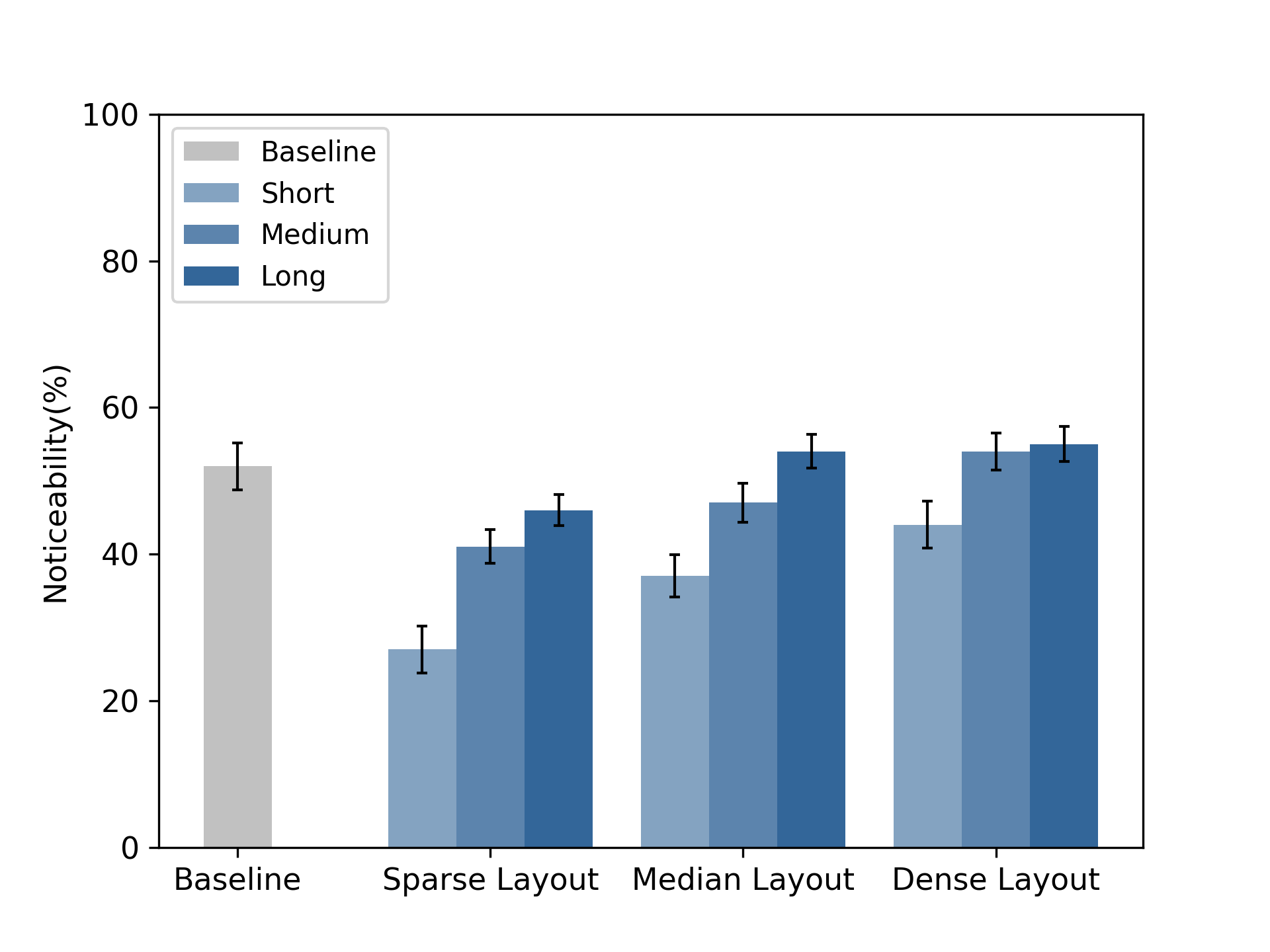}
    \caption{Noticeability results of the formative study in every condition.
    The error bars represent the standard errors.}
\end{figure}

\subsection{Results}
\label{section:study1results}

We first conducted Shapiro-Wilk tests on the noticeability results which showed that all 10 conditions followed a normal distribution, requiring no correction.
We then conducted Repeated-Measures ANOVA with Bonferroni-corrected post hoc T-tests on the results.
The average response time to visual stimuli was 327 ms (SD = 168 ms), indicating that participants were actively engaged in both tasks.

\begin{figure*}
    \centering
    \includegraphics[width=0.9\linewidth]{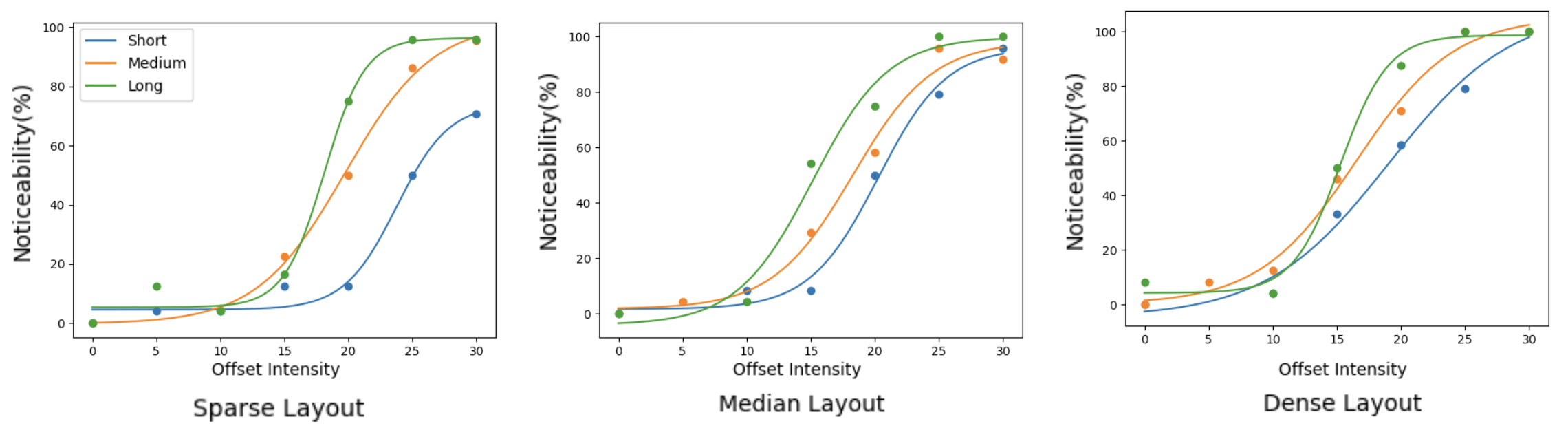}
    \caption{Psychometric functions of the noticeability in each condition.}
    \label{fig:psychometric}
\end{figure*}

\textit{With visual stimuli, participants noticed the redirection significantly less than without visual stimuli.}
We conducted a one-factor ANOVA between the baseline and the averaged nine other conditions.
Our statistical analysis showed that participants detected the redirection significantly $(F_{(1,15)}=5.90,~p=0.03)$ less frequently when they were exposed to the visual stimuli $(M=0.43,~SD=0.13)$ compared to none visual stimuli $(M=0.51,~SD=0.08)$.
These results confirm that the noticeability of redirection was reduced when visual stimuli were presented and further validate the design of our study.

We then evaluated whether participants' physical movements were effectively redirected under both noticed and unnoticed conditions. 
We divided all trials into two categories based on the participants' response to the redirection (\textit{noticed} or \textit{unnoticed}).
We then analyzed the lengths of participants' virtual and physical hand trajectories within these two categories.
The \textbf{physical trajectory length} refers to the ratio of the participant's physical hand movement trajectory length to the distance between the starting and ending pose.
Similarly, the \textbf{virtual trajectory length} refers to the participants' virtual hand movement trajectory length to the distance between the starting and ending pose.
In the unnoticed condition, participants' physical movement trajectories were significantly shorter than their virtual ones: \textit{Physical} $(AVG = 1.16,~SD = 0.04)$ and \textit{Virtual} $(AVG = 1.24,~SD = 0.05,~t(15) = 3.64,~p < 0.05)$.
Similarly, in the noticed condition, participants' physical movement trajectories were still significantly shorter than their virtual ones: \textit{Physical} $(AVG = 1.17,~SD = 0.04)$ and \textit{Virtual} $(AVG = 1.33,~SD = 0.04,~t(15) = 4.88,~p < 0.01)$.
These results suggest that participants' physical movements were successfully redirected,regardless of whether they noticed the redirection.

\begin{figure*}
    \centering
    \includegraphics[width=0.9\linewidth]{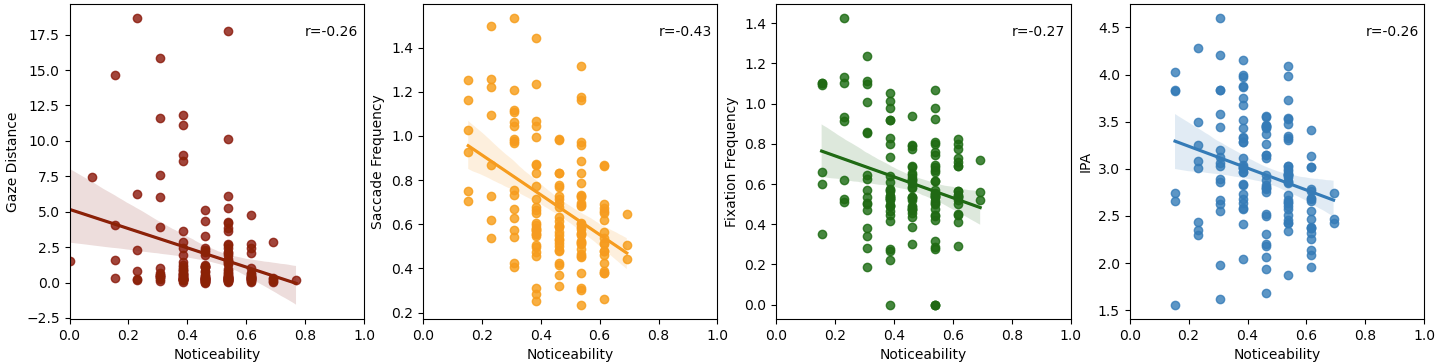}
    \caption{The regression results between gaze distance, saccade frequency, fixation frequency, IPA, and noticeability are presented. 
    The correlation coefficients are indicated in the top right corner.}
    \label{figure:correlation}
\end{figure*}

We further analyzed the gaze behaviors (gaze location, saccades, fixation, and pupil activity) based on the recorded gaze data. 
Specifically, we computed the average gaze distance relative to the virtual hand, saccade and fixation frequencies, and Index of Pupil Activity (IPA) in each condition. 
We then calculated the correlation coefficients between these gaze behaviors and the noticeability results.
As shown in~\autoref{figure:correlation}, the results revealed significant correlations: gaze distance ($r = -0.26, p = 0.001$), saccade ($r = -0.43, p < 0.001$), fixation ($r = -0.27, p < 0.001$), and IPA ($r = -0.26, p = 0.001$). 
These results suggest that all the examined gaze behaviors exhibit a negative relationship with noticeability, with gaze saccades showing a stronger effect compared to the others. 
This may be because rapid saccadic movements often indicate high cognitive load or attentional shifts, making them more directly and negatively associated with the noticing of redirection. 
In contrast, fixation frequency does not inherently reflect cognitive load, although fixation duration might serve as a useful indicator.
Regarding gaze distance to the virtual hand, participants likely shifted their gaze between the virtual body and the stimuli, making it less consistently related to noticeability. 
For IPA, its design as a long-term estimator of cognitive load may render it less sensitive to subtle or transient changes in cognitive load caused by visual stimuli.
Overall, while each of these gaze behaviors responds to visual stimuli in distinct ways, they all show promise as predictors of noticeability.


\section{Data Collection}
\label{section:datacollection}
After confirming that visual stimuli influence the noticeability of redirection, we conducted another user study using the same dual-task design to gather more data for developing a prediction model. 
This model aims to estimate noticeability based on users' gaze behavior.


\subsection{Design}

To collect the noticeability results more accurately, we measured the noticeability of each redirection magnitude repeatedly for each participant, and tested on less redirection magnitude levels.
In future work, we consider it important to extend the experiments to include a wider range of redirection magnitude.s 
As per prior work that investigated the impact of redireciton magnitude on noticeability, we chose 20 degrees as the tested magnitude, as the reported noticeability rate was around 75\% without visual stimuli in~\cite{li2022modeling}. 
The relatively high rate allowed us to detect the impact of visual stimuli effectively.
We randomly selected horizontal or vertical as the redirection direction.

\begin{figure}[!htbp]
    \centering
    \includegraphics[width=0.9\columnwidth]{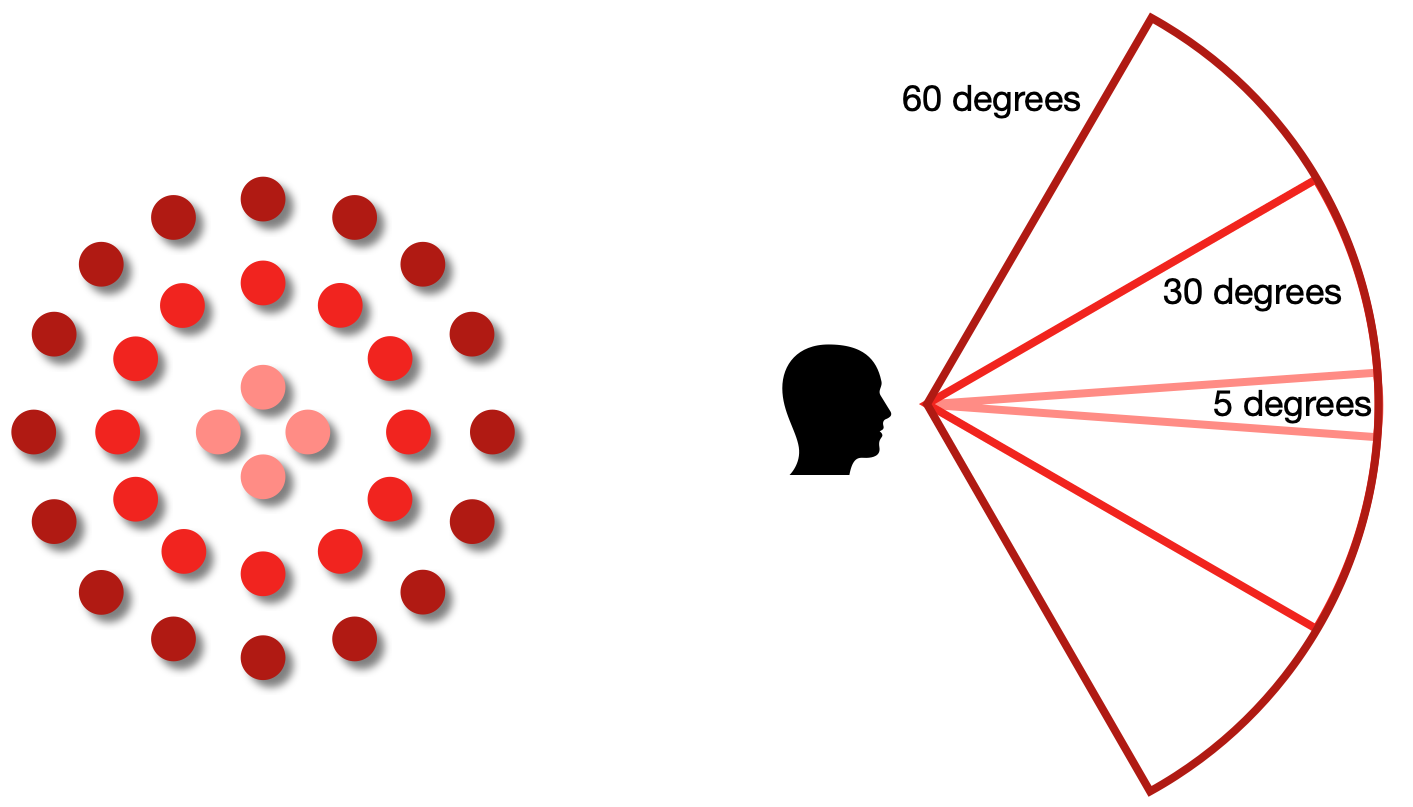}
    \caption{The possible locations of the visual stimuli in the data collection study.
    The locations are divided into three conditions: Central (5 degrees), Near Peripheral (30 degrees), and Mid Peripheral (60 degrees), based on the angular distance to the user's head direction.}
    \label{figure:datacollectionlayout}
\end{figure}

We adopted the same dual-task design with a yes/no paradigm detailed in \autoref{section:methodology}.
As our formative study results showed, the visual stimuli with medium duration ($(M=0.46,~SD=0.07)$) did not yield statistically significant differences in terms of noticeability compared to the stimuli with long duration ($(M=0.50,~SD=0.07),~(t(15)=-0.39,~p>0.05)$ ).
Therefore, we excluded the medium condition and only selected the short (0.2s) and long duration (2s) to control the intensity in this study.
To further control the position of visual stimuli within participants' visual field, we displayed the stimuli at central vision (5 degrees from the central point of vision), near peripheral vision (30 degrees), and mid peripheral vision (60 degrees), illustrated in \autoref{figure:datacollectionlayout} and as in prior research~\cite{grosvenor2007primary, gutwin2017peripheral}.
Therefore, we had $2~\times~3~=~6$ conditions, named as \textbf{CS} (central layout with short duration), \textbf{CL} (central layout with long duration), \textbf{NS} (near peripheral layout with short duration), \textbf{NL} (near peripheral layout with long duration), \textbf{MS} (mid peripheral layout with short duration), and \textbf{ML} (mid peripheral layout with long duration), and we used a Latin square to counterbalance them.
Each participant completed $(2~durations~\times~3~layouts)~\times~24$ measurements $=~144$ trials in total. 

\subsection{Apparatus \& Procedure}
The apparatus and procedure were almost identical to those of our formative study (\autoref{section:formativestudy}). 
We recorded the position and orientation of hand, elbow, shoulder, gaze, and pupil dilation with a sample rate of 60 Hz.
All gaze data was recorded with the HTC Viveo pro built-in gaze tracker.
After the warm-up session, participants took 2-minute breaks after every two sessions to reduce fatigue.
The study lasted around 40 minutes and each participant was compensated with \$15 USD.

\subsection{Participants}
We recruited 12 participants (5 females, 7 males) from a local university.
The participants had an average age of 22.91 years $(SD=1.90)$. All were trichromats and right-handed. 
Participants' self-reported their familiarity with VR at an average of 3.17 $(SD=1.27)$ on a 7-point Likert scale from 1 (not at all familiar) to 7 (very familiar).

\subsection{Summary of data statistics}

\begin{figure}[!htbp]
    \centering
    \includegraphics[width=0.9\columnwidth]{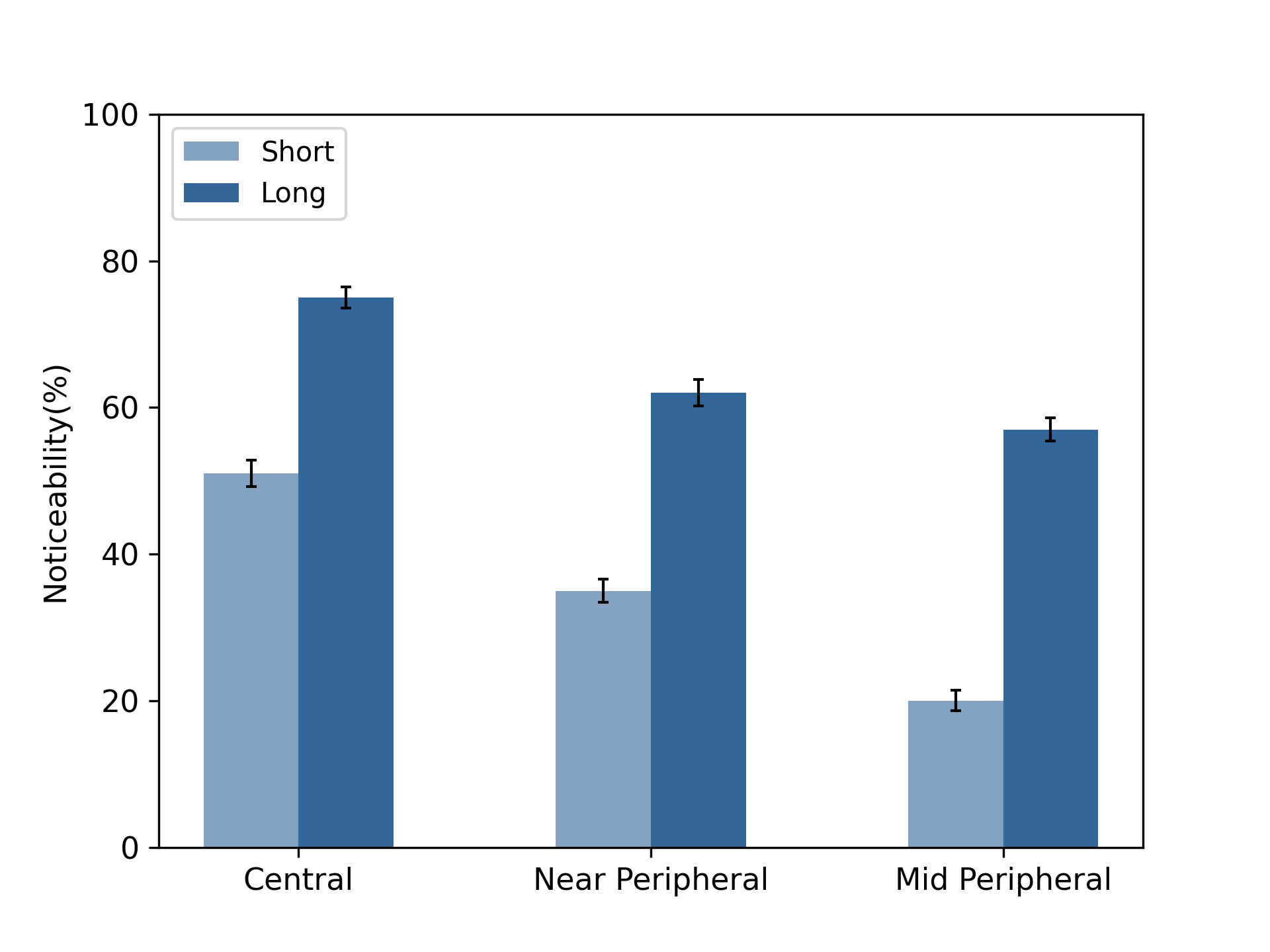}
    \caption{Noticeability results of the data collection study in each condition. The error bars represent the standard errors.}
    \label{figure:datacollectionresult}
\end{figure}

In total, we collected 1728 responses.
To estimate the noticeability, we calculated the ratio of trials in which participants reported noticing the redirection to the total number of trials for each session and participant.
As shown in \autoref{figure:datacollectionresult}, the noticeability result differed across the visual stimuli's duration and layout.
The noticeability results ranged from 16.7\% to 79.2\% with an average of 50.1\% and a standard deviation of 18.9\%.
The maximum and minimum indicate that we controlled the noticeability with the visual stimuli's duration and layout successfully.
Additionally, the high standard deviation suggest a high variability across conditions, which is beneficial for training a model to predict the influence of visual stimuli on noticeability.

To verify that participants' physical movements were effectively redirected, we analyzed participants virtual and physical trajectory lengths as defined in~\autoref{section:study1results}.
In the unnoticed condition, participants' physical movement trajectories were significantly shorter than their virtual ones: \textit{Physical} $(AVG = 1.14,~SD = 0.04)$ and \textit{Virtual} $(AVG = 1.20,~SD = 0.04,~t(11) = 3.97,~p < 0.05)$.
In the noticed condition, participants' physical movement trajectories were also significantly shorter than their virtual ones: \textit{Physical} $(AVG = 1.14,~SD = 0.04)$ and \textit{Virtual} $(AVG = 1.27,~SD = 0.05,~t(11) = 5.38,~p < 0.01)$.
These results indicated that participants' physical movements were successfully redirected.

\section{Implementation}
\label{section:implementation}

In this section, we investigated the best combination of gaze behavior features to predict the noticeability of redirection.
We described participants' gaze behaviors with pupil activity, gaze angular distance to users' hands, eye saccade, and fixation.
For each category of participants' gaze behavioral data, we systematically examine different feature combinations to identify those that most accurately characterize the participants' visual responses.
We then develop a regression model to explain the relationship between the selected gaze behavioral features and the noticeability of redirection.


\subsection{Gaze behavioral patterns}
\label{section:featureselection}

Results in \autoref{section:formativestudy} showed that the users' gaze behavioral data was correlated to the noticeability.
We divided the gaze behavioral data into four categories:

\begin{itemize}
    \item \textit{Index of Pupillary Activity}: 
    Index of Pupillary Activity (IPA) has been used to reflect users' cognitive load by analyzing the change of users' pupil dilation~\cite{duchowski2018index, lindlbauer2019context}. 
    While visual stimuli were presented, users' cognitive load might be inadvertently affected and could therefore impact the noticeability results.
    \item \textit{Gaze angular distance to elbow and hand}:  
    We calculated the vector starting from the user's eye to the elbow and hand joint. 
    We then calculated the angular distance between this vector and the gaze vector.
    These metrics reflect whether the participant was looking at the primary task or attracted by the visual stimuli.
    We decided not to calculate the distance between the focus point and the visual stimuli, as in real-world use cases, there is no single visual stimuli but only complicated ones, which make it hard to compute this distance.
    \item \textit{Eye saccade frequency, duration and interval}:
    Eye saccade is a rapid eye movement that shifts the eye from one area to another.
    We leveraged the detection algorithm from \citeauthor{pymovements} to detect the saccade frequency and duration~\cite{pymovements}. 
    The saccade frequency and duration indicate how often and how quickly users shift their eye gaze separately.
    The saccade interval suggests the temporal distribution, which indicates whether the saccades are uniformly distributed across the session.
    \item \textit{Eye fixation frequency, duration and interval}:
    Eye fixations represent when eyes stop scanning the scene and hold the foveal vision on an object of interest. 
    We also used the frequency, duration, and interval of eye fixation to indicate how often and how long users stared at a place and the temporal distribution of fixation.
\end{itemize}

\subsection{Regression model}

To better represent the previous gaze behavioral patterns, we computed \textit{mean, standard deviation, median, maximum and minimum} of IPA and gaze angular distance and combined them with the eye saccade and fixation features.
Therefore, we had 3 behaviors(IPA, gaze angular distance to hand, gaze angular distance to elbow) $\times$ 5 features (mean, standard deviation, median, maximum and minimum) $+$ 3 saccade (saccade frequency, duration and interval) $+$ 3 fixation (fixation frequency, duration and interval) $=~21$ features in total.
However, the search space to determine the combination of features that provides the highest predictive power for noticeability includes as many as $\sum_{i=1}^{21}\frac{21!}{i!(21-i!)}~=~2^{21}-1$ conditions, which means that a grid search is not practical.
Therefore, we adopted a similar method as \cite{maslych2023effective} to select the features.
We first selected the best combination of features within each category and then searched the combination of these categories iteratively to figure out the best combination.

In this process, we used Support Vector Regression (SVR) from scikit-learn package~\footnote{https://scikit-learn.org/stable/modules/generated/sklearn.svm.SVR.html} as the benchmark model since SVR has a stable performance on various data.
The SVR model took the selected features as input, then output a probability ranging from 0 to 1 as the predicted noticeability.
We leveraged the leave-one-user-out cross-validation in the test and the mean squared error (MSE) as the metric.

\autoref{table:featureselection} lists the best combination of features within each of four categories.
Among them, the selected combination in the \textit{gaze angular distance} achieved the best performance, while the other features also demonstrated the potential for predicting noticeability.
Therefore, we combined the features from different categories and further tested them.

\begin{table*}[htb]
  \centering
  \small
    \begin{tabular}[width=\columnwidth]{ccc}
    \toprule
    \textbf{Category} & \textbf{Best combination} & \textbf{MSE}\\
    \midrule
    IPA & mean, maximum, minimum & 0.039 (0.013) \\
    Gaze angular distance & \makecell{mean(hand), std(hand), median(hand) \\ mean(elbow), std(elbow), maximum(elbow)} & 0.017(0.008) \\
    Eye saccade & frequency, duration, interval & 0.027(0.009) \\
    Eye fixation & frequency, duration & 0.040 (0.012) \\
    \bottomrule
    \end{tabular}
    \caption{The best feature combination of each category and the prediction performance. The prediction performance is presented as the average (standard deviation) of MSE.}
    \label{table:featureselection}
\end{table*}

\begin{figure}[t]
    \centering
    \includegraphics[width=0.9\columnwidth]{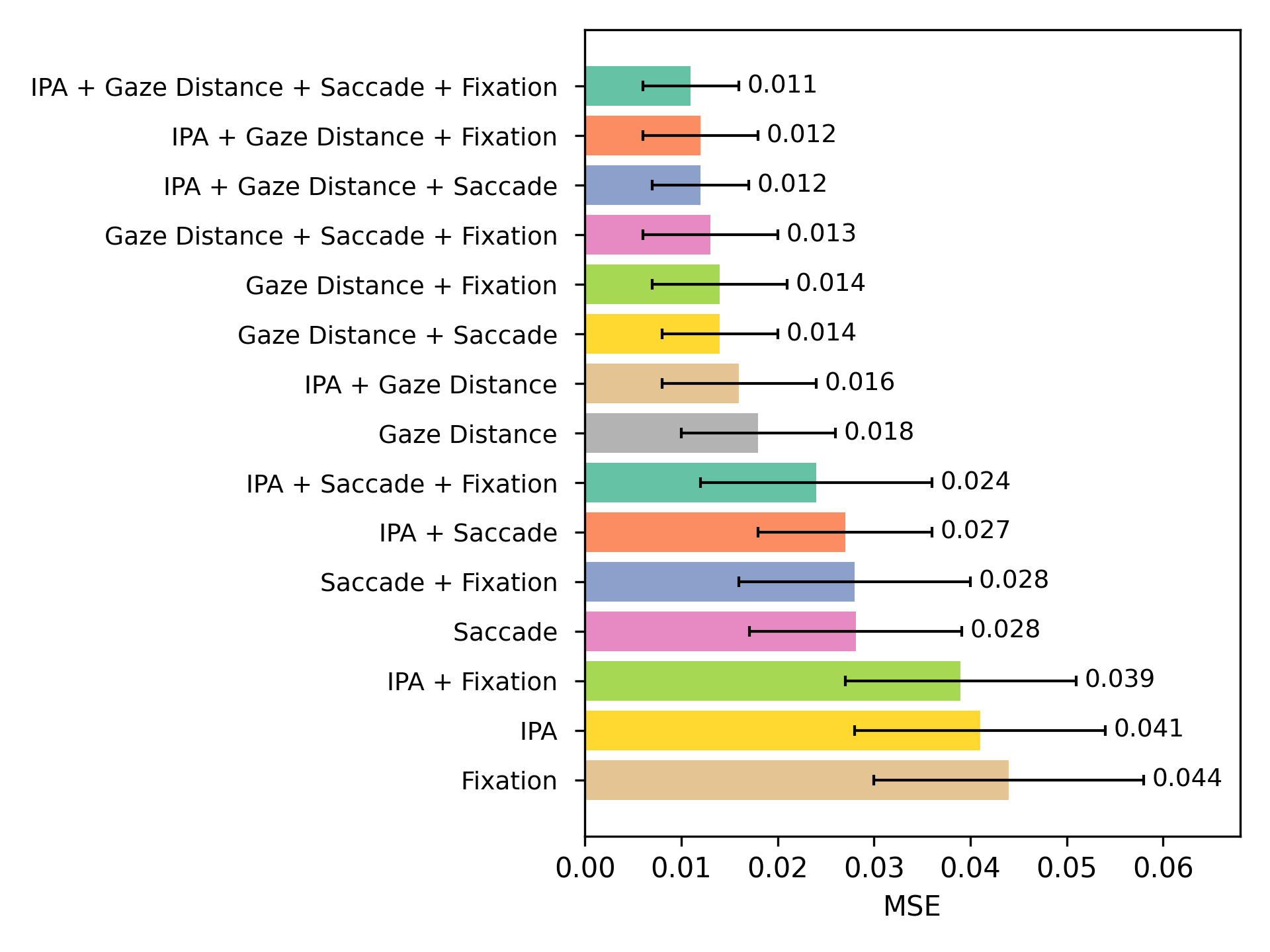}
    \caption{The regression error of all combinations of the feature category. The error bars denote the standard deviations.}
    \label{figure:featureselection}
\end{figure}

We then tested the regression error of all combinations of the feature category with leave-one-user-out cross-validation.
For each feature combination, we filtered the data with it and then fitted a model with 11 participants' data and tested it on the one remaining participant's data.
After repeating this 12 times, we determined the overall regression error for one feature combination.
\autoref{figure:featureselection} illustrates the regression error of all 15 feature combinations.
The results demonstrate that combining all these four category features achieves the best performance with an MSE of 0.011.

To further understand the best feature combination across the users, we also explored the best feature set for each test user in the leave-one-user-out cross-validation process.
For each test user, we trained a model with each feature combination and selected the best one.
The results showed that for 7 out of the 12 participants, the best feature set was the combination of all four feature categories.
For 3 of the 12 participants, the best set was the combination of IPA, Gaze Angular Distance and Fixation and for the other 2 of the 12 participants, the best set contained IPA, Gaze Angular Distance and Saccade.
The results suggest that each feature captures distinct aspects of gaze behavior that contribute to predicting noticeability. 
Although the gaze angular distance showed a lower correlation with noticeability compared to saccades in \autoref{section:formativestudy}, it performs as the most powerful feature for predicting noticeability. 
This may due to the fact that in~\autoref{section:formativestudy}, we only considered the mean distance, whereas in this study, we included additional numerical features, which could provide more informative insights than the mean alone.
As for eye gaze saccade, it also contributes significantly to the prediction, aligning with the correlation results in~\autoref{section:formativestudy}, as it indicates users' visual focus shift and cognitive activity.
While IPA and fixation also have the potential to predict noticeability, their prediction accuracy is lower compared to the other features. 
This could be because they reflect more general cognitive activity and engagement, rather than specific responses to visual stimuli.
However, combining these features allows us to capture both where users are looking at and the dynamic shifts in focus, which together indicate the noticeability of redirection.

\begin{figure*}[t]
    \centering
    \begin{subfigure}{0.47\linewidth}
        \includegraphics[width=\columnwidth]{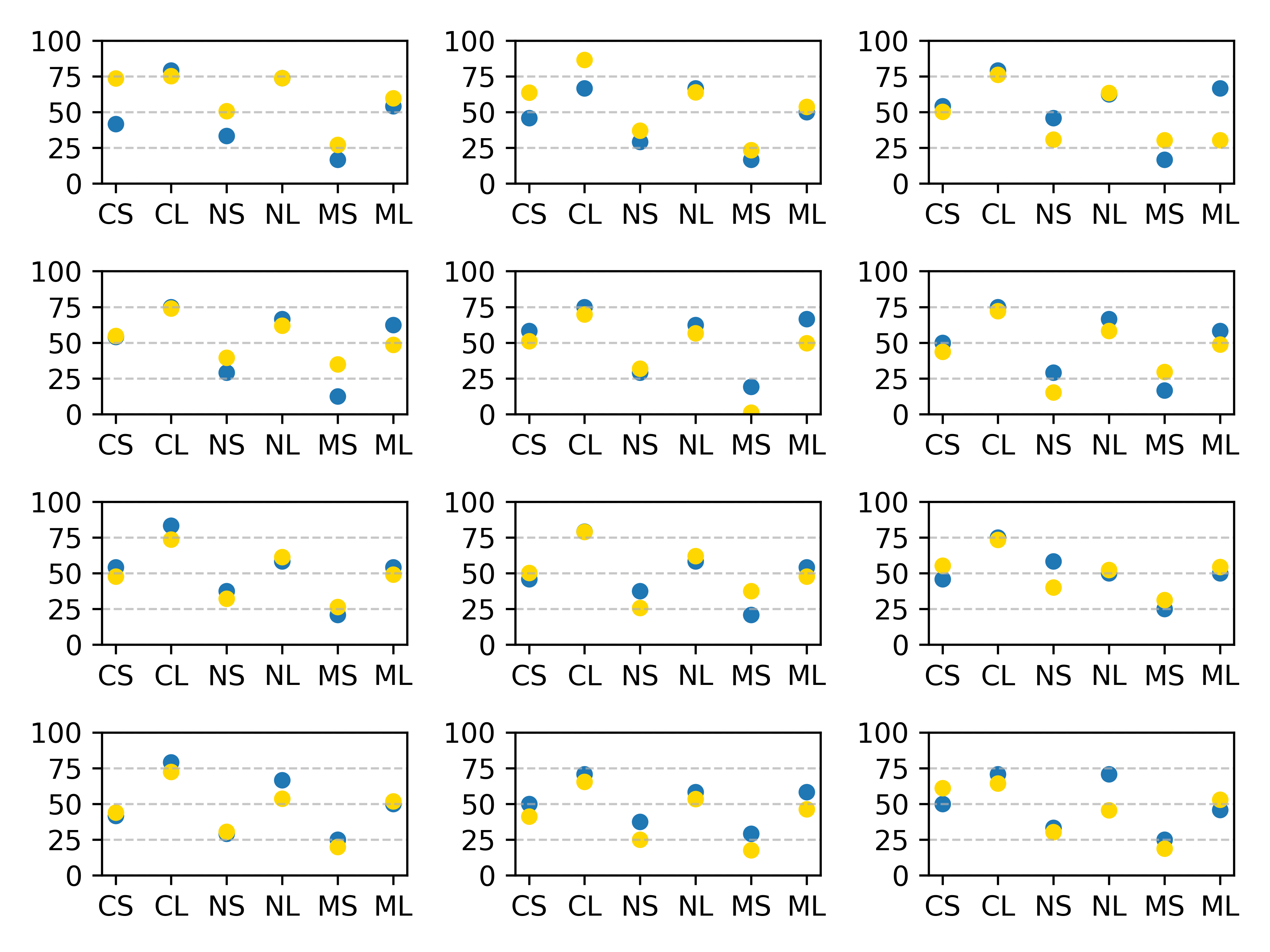}
        \caption{}
        \label{figure:predictionperformance_peruser}
    \end{subfigure}
    \centering
    \begin{subfigure}{0.47\linewidth}
        \includegraphics[width=\columnwidth]{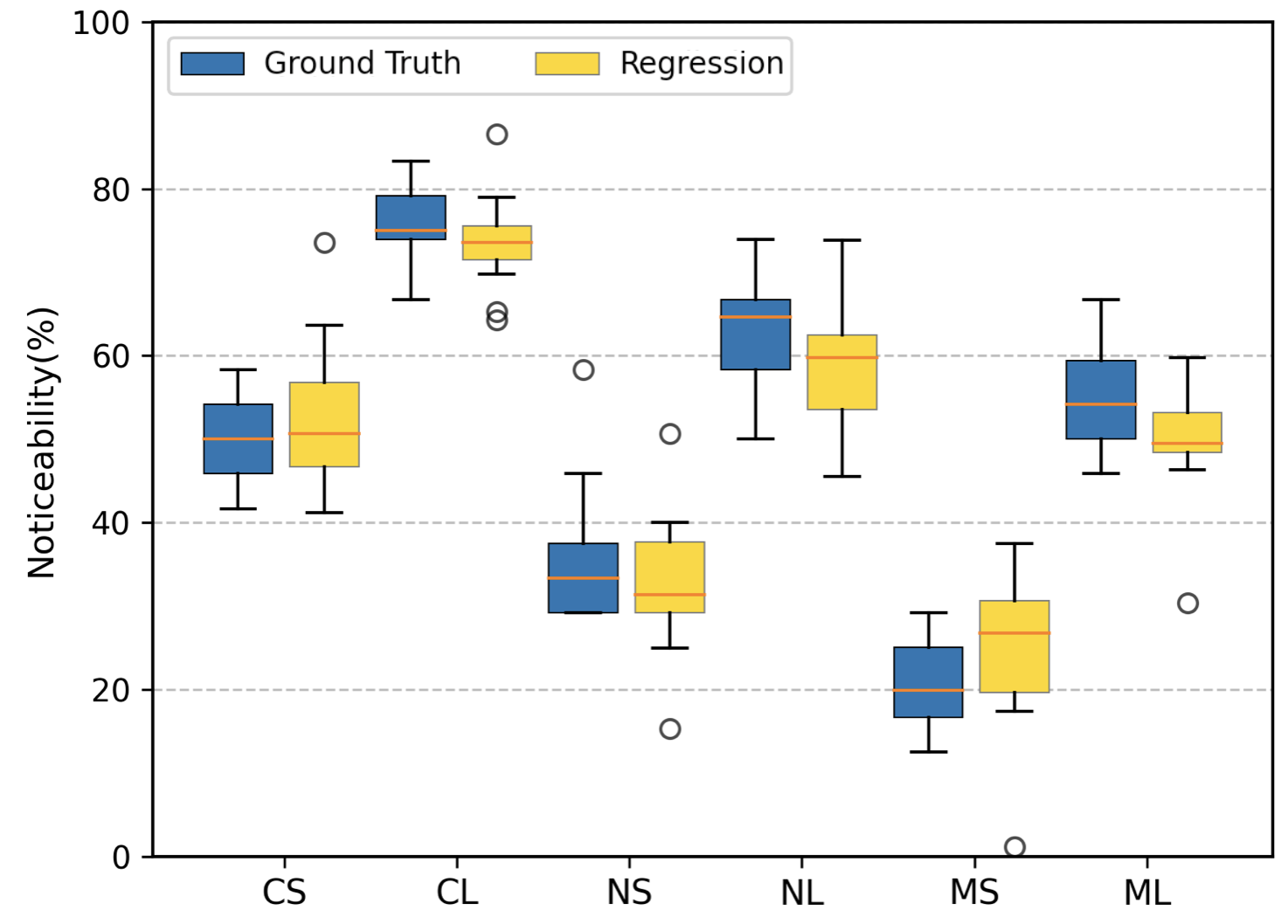}
        \caption{}
        \label{figure:predictionperformance_average}
    \end{subfigure}
    \caption{(a) illustrates the regression results in each condition for each user. (b) illustrates the regression results with leave-one-user-out cross-validation.}
    \label{figure:predictionperformance}
\end{figure*}

To further investigate if the selected features could model noticeability, we analyzed the regression error for each individual data point in a per user manner. 
As shown in \autoref{figure:predictionperformance_peruser}, the outputs from our model preserved the relative order of noticeability across the six conditions in 90.3\% data points.
The fitted noticeability in various conditions mostly remained in the range of the ground truth, while most errors came from the two most similar conditions (\textbf{CS} and \textbf{NL}). 
Furthermore, \autoref{figure:predictionperformance_average} illustrates the noticeability average and standard deviation of the data collection results and our model's output.
Our model's output average approximates the participant's results while simultaneously exhibiting a lower standard deviation.
This could be due to the inherent noise introduced from estimating the noticeability using the frequency of participants who reported the noticing of redirection in the study.

\subsection{Classficiation model}
\label{section:classfication_model}

In our studies, noticeability was measured by the frequency with which participants detected redirection during the trials. Based on this, we developed a regression model that outputs the probability of noticing the redirection as a float value between 0 and 1. While this probability effectively indicates how likely users are to notice the redirection, a classification model providing a simple yes/no result could offer greater practical utility.
To explore this, we trained a classification model by applying various thresholds to the noticeability results and categorizing it into distinct classes. 

\begin{description}
\item[Binary model]
We applied a threshold of 0.5 to transform the collected noticeability results into binary labels: Unnoticeable $(\leq 0.5)$ and Noticeable $(> 0.5)$.
With these, we trained a Support Vector Machine (SVM) classification model with the same features selected in \autoref{section:featureselection}; this model achieved an accuracy of 0.9174 $(SD=0.1126)$ and an F1-score of 0.8968 $(SD=0.1342)$ with leave-one-user-out cross-validation on our collected dataset.
\item[Three-class model]
Then we divided the noticeability into three categories with two thresholds: Low Noticeability $(\leq 0.4)$, Medium Noticeability $(0.4 <$ noticeability $\leq 0.7)$, and High Noticeability $(>0.7)$.
With the same SVM classification model and selected feature, our re-trained model achieved an accuracy of 0.8562 $(SD=0.1240)$ and an F1-score of 0.8478 $(SD=0.1276)$.
To be noted, the prediction accuracy was affected by how we converted the noticeability value to separate labels and might increase with fine-tuned features tailored to the classification task.
This indicates that the selected features from the gaze behavioral pattern have the potential to predict the noticeability as separate categories.
\end{description}

\section{Evaluation: Extend to More Visual Stimuli}
\label{section:evaluation}

To evaluate the performance and extendability of our regression model, we tested our model with the same dual-task experiments and yes/no procedure, but with new visual stimuli and new participants.
We tested scale- and color-based visual stimuli, which were not included in the training set of the proposed model.
As the model only takes the user's gaze behavioral data as input without any prior knowledge about the visual stimuli, we aimed to also investigate whether the gaze behavioral patterns of users remain consistent with different types of stimuli.

\subsection{Design}

\begin{figure}[htbp]
    \centering
    \includegraphics[width=0.9\columnwidth]{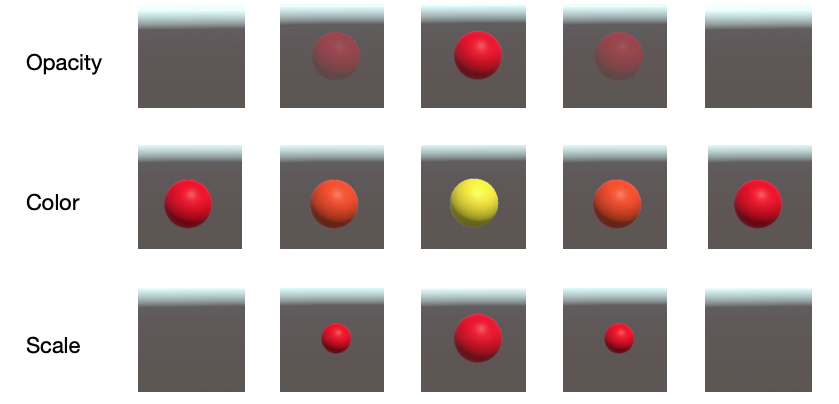}
    \caption{Demonstrations of the three visual effects (opacity, color, scale) in the formative study and the evaluation study.}
    \label{figure:visualeffects}
\end{figure}

We designed two new visual stimuli that the participants were required to monitor and report in the secondary task in this study, as shown in \autoref{figure:visualeffects}.
In a scale-based animation, a red ball increased from an invisible small scale to the normal scale of the same as in the opacity stimuli and reset to the invisible small scale.
As for the color-based animation, the red ball would change from red to yellow and reset to red by altering the hue in the HSV color space.
We note that in the opacity- and scale-based animations, the red ball started at an invisible state and we could adjust the ball's location while it was invisible to users.
However, in the case of the color-based animation, the red ball remained consistently visible.
Thus, we adjusted the ball's location when the color animation finished and waited a random time interval before the start of the color visual effect.
We asked participants to report as soon as the color changed to yellow instead of the location change.

For both visual stimuli, we used the duration and layout of the visual stimuli to control their influence on the noticeability.
Therefore, we had $2~\times~3~=~6$ conditions with the order counterbalanced by a Latin square.
However, to avoid the influence of fatigue, we adopted a between-subject study design for the two visual stimuli conditions, where each participant only tested either color- or scale-based stimuli.
In this way, each participant completed $1$ visual effect $\times~(2~durations~\times~3~layouts)~\times~24$ measurements $=~144$ trials in total. 

We calculated the selected features based on the participant's gaze data, as described in \autoref{section:featureselection}, and leveraged our SVR model to output the noticeability.

\subsection{Apparatus \& Procedure}
The apparatus and procedure were identical to those of our data collection study (\autoref{section:datacollection}). 
The study lasted around 40 minutes and each participant was compensated with \$15 USD.

\subsection{Participants}

We recruited 24 new participants from a local university for the study and divided them into two groups randomly.
For the color-based visual effects, 12 participants (5 females, 7 males, average age of 22.83 years with $SD=1.27$) were allocated to participate in the scale-based effect study.
These participants reported their familiarity with VR as an average of 3.83 $(SD=1.75)$ on a 7-point Likert-type scale from 1 (not at all familiar) to 7 (very familiar).
For the scale-based visual effects, we had another 12 participants (6 females, 6 males) with an average age of 22.75 $(SD=1.76)$ and a self-reported familiarity with VR of an average of 3.25 $(SD=1.22)$.
All participants were right-handed.

\subsection{Result}

\begin{figure*}[htbp]
    \centering
    \begin{subfigure}{0.47\linewidth}
        \includegraphics[width=0.8\columnwidth]{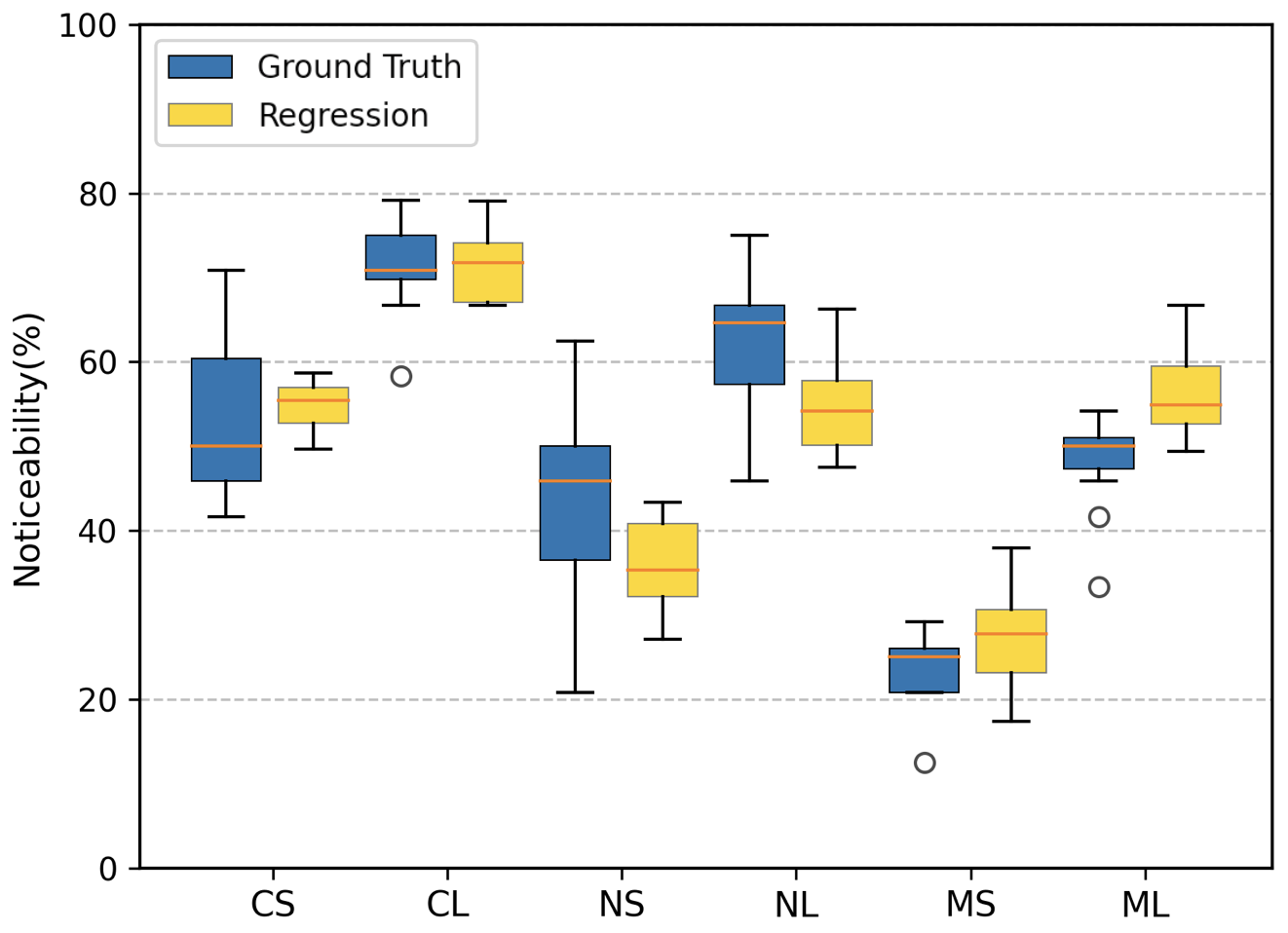}
        \caption{Study results for color-based visual effects.}
        \label{figure:evaluationresult_clr}
    \end{subfigure}
    \centering
    \begin{subfigure}{0.47\linewidth}
        \includegraphics[width=0.8\columnwidth]{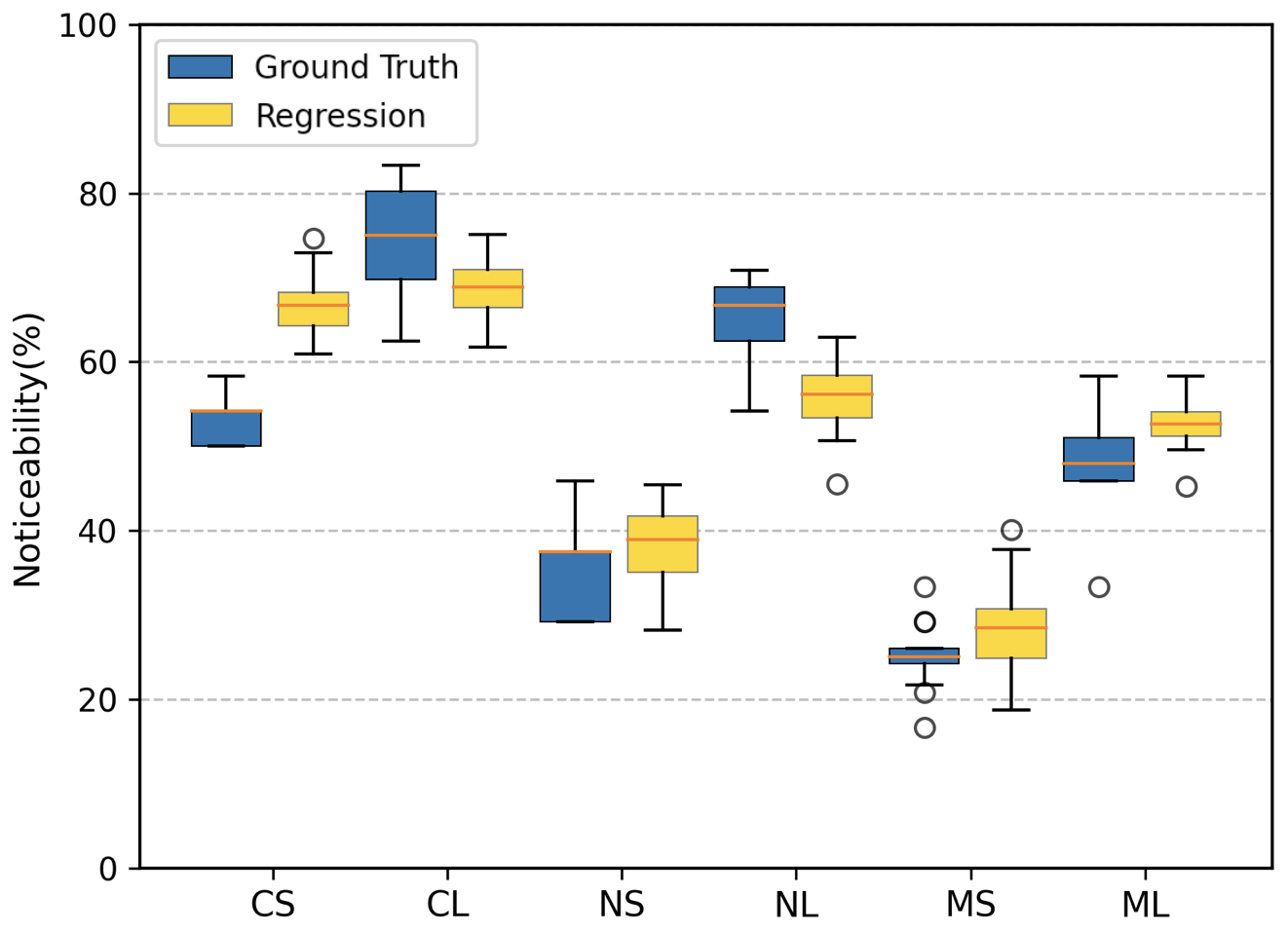}
        \caption{Study results for scale-based visual effects.}
        \label{figure:evaluationresult_scl}
    \end{subfigure}
    \caption{Study results of the model trained with opacity-based visual effects, tested on the color- and scale-based visual effects.}
    \label{figure:evaluationresult}
\end{figure*}

As shown in \autoref{figure:evaluationresult}, the participants' noticeability of arm redirection with color-based visual stimuli varied from 12\% to 79\% in six conditions with an average of 49\% and a standard deviation of 17\%.
Similarly, the noticeability with scale-based visual stimuli ranged from 17\% to 83\% with an average of 51\% and a standard deviation of 18\%.
This suggested that the noticeability of redirection was effectively affected by the duration and layout of the color- and scale-based visual stimuli.

We further analyzed the virtual and physical trajectory lengths under different noticeability conditions to confirm whether participants' physical movements were effectively redirected, as the same in~\autoref{section:study1results}.
In the unnoticed condition, participants' physical movement trajectories were significantly shorter than their virtual trajectories: \textit{Physical} $(AVG = 1.17,~SD = 0.05)$ and \textit{Virtual} $(AVG = 1.27,~SD = 0.05,~t(11) = 2.48,~p < 0.05)$.
Similarly, in the noticed condition, participants' physical movement trajectories were also shorter than their virtual trajectories: \textit{Physical} $(AVG = 1.15,~SD = 0.04)$ and \textit{Virtual} $(AVG = 1.28,~SD = 0.04,~t(11) = 4.88,~p < 0.01)$.

\textit{Our regression model could accurately compute the noticeability with new color- and scale-based visual stimuli.}
We calculated the selected features with the recorded eye gaze behavior data.
Our model takes the selected features as input and outputs the noticeability in different conditions with leave-one-user-out cross-validation.
For this study, our model achieved an MSE of 0.014 (SD = 0.006) and 0.012 (SD = 0.005) for noticeability with the color- and scale-based visual stimuli separately.
Compared to the 0.011 MSE result of the opacity visual stimuli, these results indicate that our model has the potential to compute noticeability accurately under new visual stimuli with new participants.
This indicates that although our model was built with the data collected from a limited number of participants, the elicited eye behavioral patterns could be generalized to different visual effects and new users.
We note that the two tested stimuli never appeared in the training dataset of our model, and the color-based stimuli even apply a paradigm of being always visible different from the opacity-based stimuli that the model was trained with.
Our results show that our model has the potential to be applied to scenarios with different visual stimuli, as long as the gaze behavioral patterns of users are consistent across scenarios.

\textit{Participants' gaze behavioral patterns are consistent across three conditions of visual stimuli.}
To further explore if participants exhibit similar gaze behavioral patterns when testing different visual stimuli, we conducted a technical evaluation with the data recorded in the previous study.
We leveraged the selected features to train a regression model with the data from one of the three visual stimuli (opacity, color, and scale) and tested the model on the data from the other two stimuli.
As shown in \autoref{table:evaluationresult}, the regression model of color- and scale-based visual stimuli also achieved a comparable performance when computing the noticeability under the same visual stimuli.
While all three regression models achieved the best performance with the test data from the same visual effect, they also proved the ability to compute the noticeability under other two visual stimuli.
This suggests that the gaze behavioral patterns were consistent across visual stimuli, and can be used to compute the noticeability of redirection.

\begin{table}[htb]
  \centering
  \small
    \begin{tabular}[width=\columnwidth]{cccc}
    \toprule
    \textbf{Train set} & \textbf{MSE of Opacity} & \textbf{MSE of Color} & \textbf{MSE of Scale} \\
    \midrule
    Opacity & 0.011(0.005) & 0.014(0.006) & 0.012(0.005) \\
    Color & 0.018(0.014) & 0.011 (0.005) & 0.015(0.009) \\
    Scale & 0.016(0.009) & 0.022(0.014) & 0.013(0.007) \\
    \bottomrule
    \end{tabular}
    \caption{The regression performance of training the model with data under one visual stimulus and testing on the data from all three visual stimuli. The results are presented as the average (standard deviation) of MSE.}
    \label{table:evaluationresult}
\end{table}

\section{Towards real-world use cases}

While the previous study results suggest that our proposed model could effectively compute the noticeability of redirection under various basic visual stimuli (transparency-, color- and scale-based), we aimed to explore how the model could be used in real-world scenarios.
To showcase the potential benefits of our model in practical use cases, we implemented \textbf{an adaptive redirection technique} and developed \textbf{two real-world applications} to demonstrate its generalizability and usability.
We also performed a proof-of-concept study to gather user feedback while interacting with the two applications and the adaptive redirection technique.



\begin{figure}[t]
    \centering
    \begin{subfigure}{0.45\columnwidth}
        \includegraphics[width=0.9\columnwidth]{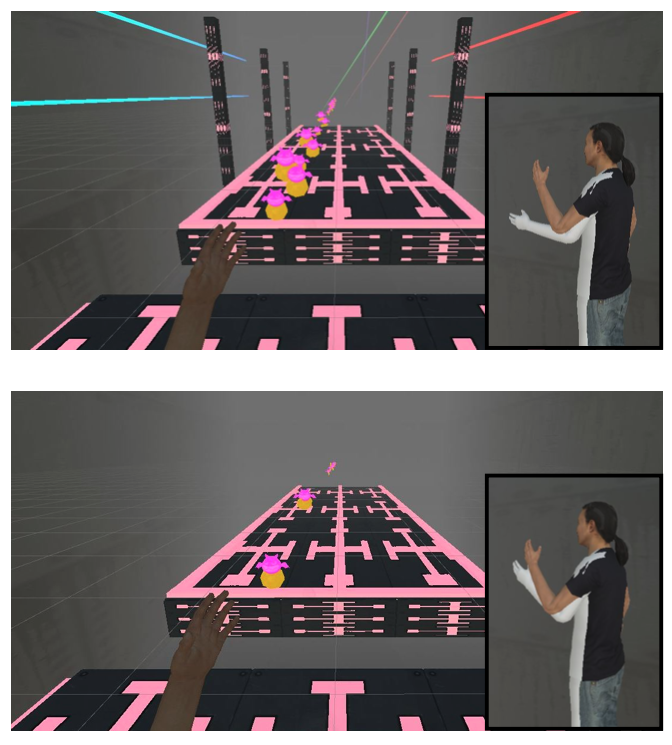}
        \caption{}
        \label{figure:application_adaptive}
    \end{subfigure}
    \centering
    \begin{subfigure}{0.45\columnwidth}
        \includegraphics[width=0.9\columnwidth]{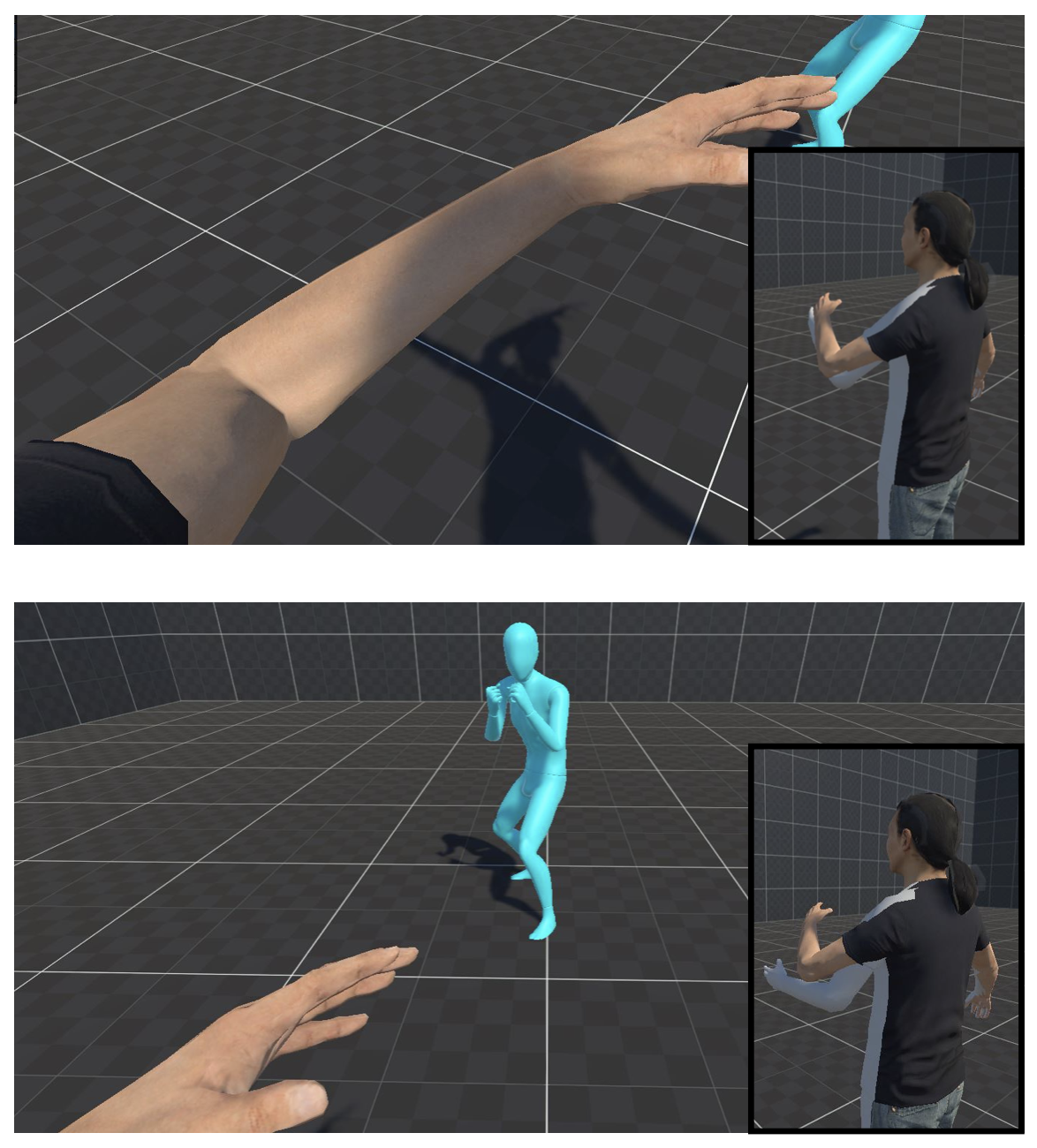}
        \caption{}
        \label{figure:opportunistic}
    \end{subfigure}
    \caption{
    We developed two real-world applications to demonstrate the capabilities of our adaptive redirection technique:
    (a) Adjusting the difficulty of VR action game: 
    In this application, mid-air coins and monsters serve as visual cues for the target poses that users are asked to perform. 
    Our adaptive redirection technique enables the system to adjust the game’s difficulty without the user noticing, ensuring a balanced and engaging experience.
    (b) Opportunistic rendering for boxing training in VR : 
    Here, users are learning boxing movements by following a blue avatar. 
    Leveraging our adaptive redirection technique, the system can simulate opportunistic rendering which reduces requirements for computation resources.}
    \label{figure:application}
\end{figure}

\subsection{Adaptive motion redirection technique}
As discussed in the Introduction (\autoref{section:introduction}) and Related Work (\autoref{section:related_work}), users in real VR applications may face complicated visual effects that can impact the noticeability of redirection movements. 
This, in turn, influences the effectiveness and overall user experience of redirection techniques.
While it is impractical to predict the specific visual effects users will encounter beforehand, content creators can only predefine a static redirection intensity, which limits the effectiveness of redirection techniques. 
To address this limitation, our proposed model enables designers to dynamically adjust the redirection during usage based on the user’s gaze behavior.

Our model computes the noticeability of redirection as a float value ranging from 0 to 1. 
With this output, we implemented an adaptive redirection technique by using the Three-class model described in \autoref{section:classfication_model}.
For each class of noticeability, we predefined corresponding redirection: 25 degrees for Low Noticeability, 15 degrees for Medium Noticeability, and 5 degrees for High Noticeability. 
When the computed noticeability falls into one of these classes, the corresponding redirection is applied.
The redirection technique initializes with a 10 degree offset. 
When a change in redirection is required according to the noticeability changes, we use linear interpolation to transition the redirection gradually over a 10-second period. 
To maintain immersion, the redirection is adjusted only when the user’s arm is in motion, since if the redirection changes while the physical arm remains static, the virtual arm will be moved and lead to break of immersion and sense of embodiment.

To be noted, this adaptive redirection technique serves as a demonstration of the usability of our proposed model. 
Designers can leverage the model's probabilistic output to create their own redirection techniques tailored to specific applications.

\subsection{Real-world applications}

\subsubsection{Adjusting the difficulty of VR action game}
Based on our adaptive redirection technique, we implemented a VR action game inspired by the VR game Beat Saber~\footnote{https://beatsaber.com/}.
In the game, users are asked to perform certain poses with their arms based on visual and musical guidance.
The game difficulty could be adjusted by redirecting the user's movement, for example, slightly amplifying their movements could make it easier and faster to achieve the targets.
Meanwhile, users need to focus on the targets to obtain sufficient information, and thus they paid less visual attention to their virtual body movements.
As shown in \autoref{figure:application_adaptive}, when the visual guidance for the target arm pose is highly detailed and draws significant attention from the user, 
the noticeability of redirection might be lower and 
the system can take the risk of applying large redirection for functional gains.
However, when the user interacts with a simpler interface and focuses mainly on their virtual arm, 
a low level redirection might be applied with the high noticeability prediction.


\subsubsection{Opportunistic rendering for boxing training in VR}
We implemented a boxing training system designed to reduce rendering computation as our second application.
Accurate motion reconstruction and rendering may require high computing power~\cite{chen2021towards}.
While users may not always focus on their virtual movements, there is a chance to apply opportunistic rendering based on the user's visual attention to save computing capability and avoid being noticed by users.
As shown in \autoref{figure:opportunistic}, the user is learning boxing poses with a virtual coach in VR.
When the user is looking at the coach and observing them performing the pose, our model may output a lower level of noticeability and thus it allows the system to update the user's movement less frequently which leads to the virtual movement has a offset with the user's physical movement and save computing resources.
While the user shifts their attention back to his arm and is going to practice the boxing poses, our model can compute that the noticeability of motion offset is higher than in the previous scenario.
Therefore, the system can allocate more resources to render the user's movement, to ensure that they can perform and learn the accurate poses in VR.
To be noted, we implemented this application as a simulation of opportunistic rendering to demonstrate the potential of our model, rather than fully implementing it and measuring the computational resources it would save.


\subsection{Proof-of-Concept study}

To further demonstrate and evaluate the how our model supports adaptive redirection techniques, we conducted a proof-of-concept evaluation study on two applications.

\subsubsection{Design}

We conducted a within-subject factorial study design, with the independent variable being the experimental conditions, including Adaptive Redirection (\textbf{AR}) and Static Redirection (\textbf{SR}).
In the VR action game, participants were tasked with performing poses that aligned with a moving target. 
The target’s appearance frequency progressively increased, starting at intervals of 2 seconds and accelerating to 0.5 seconds and the game lasted for 60 seconds.
In the boxing training application, participants engaged in a 60-second motion-learning task, attempting to replicate the movements demonstrated by a virtual coach.
For the \textbf{SR} condition, the redirection magnitude was fixed at 15 degrees, which is the same as the medium level magnitude used in the \textbf{AR} condition.
After completing the tasks in each condition, participants rated the tested conditions on physical demand ("\textit{The interaction was physically demanding}"), mental demand ("\textit{The interaction was mentally demanding and I had to concentrate a lot.}"), embodiment ("\textit{I felt as if the virtual body was my body}") and agency ("\textit{I felt like I could control the virtual body as if it was my own body}") with a 7-point Likert scale, using the questions from similar studies in prior work~\cite{peck2021avatar, feick2023investigating}.

\subsubsection{Apparatus \& Procedure}

We implemented the applications with a HTC Vive pro headset in Unity 2019, powered by an Intel Core i7 CPU and an NVIDIA GeForce RTX 2080 GPU. 
During the study, participants were equipped with three Vive Trackers affixed to their left shoulder, elbow, and waist using nylon straps.

After being introduced to the study, participants had a warm-up session to learn about the study tasks and get familiar with controlling the virtual movements.
Once they were comfortable with the virtual movements and tasks, they proceeded to experience one condition across both applications.
After completing the two applications under the first condition, participants provided their ratings before moving on to experience the second condition. 
The order of conditions and applications was counterbalanced.
The study lasted around 20 minutes, and each participant received a compensation of 10 US dollars for their participation.

\subsubsection{Participants}

We recruited 8 new participants (2 females, 6 males, average age of 25.63 with $SD=1.85$) from a local university.
These participants reported their familiarity with VR as an average of 3.75 $(SD=1.16)$ on a 7-point Likert-type scale from 1 (not at all familiar) to 7 (very familiar).

\subsubsection{Result}

\begin{figure}[t]
    \centering
    \includegraphics[width=0.9\linewidth]{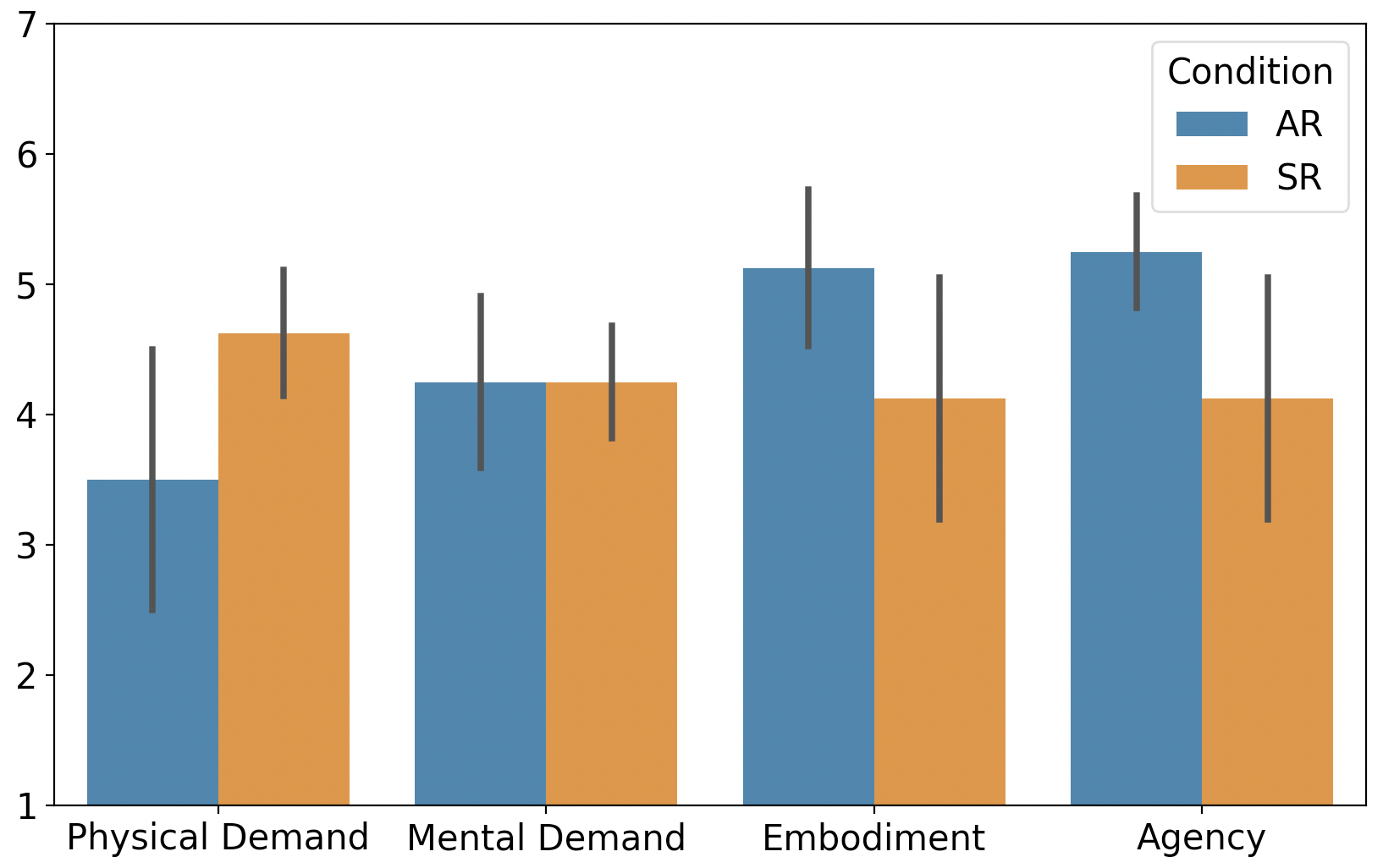}
    \caption{Proof-of-concept study results indicate that participants experienced less physical demand and a stronger sense of embodiment and agency when using the adaptive redirection technique compared to the static technique.}
    \label{figure:preliminary_result}
\end{figure}

\autoref{figure:preliminary_result} summarizes the study results.
We conducted Wilcoxon signed-rank tests to analyze the reported subjective metrics.
Participants reported lower physical demand in the adaptive redirection (\textbf{AR}) condition ($M=3.50, SD=1.00$) compared to the static redirection (\textbf{SR}) condition ($M=4.50, SD=0.50$, $W=2.00, p<0.05$). 
This is due to the larger redirection allowed in \textbf{AR} when visual stimuli were intense, reducing the need for extensive physical movement.
Despite the adaptive nature of \textbf{AR}, participants did not perceive a higher mental demand ($M=4.25, SD=0.60$) compared to \textbf{SR} ($M=4.25, SD=0.43$, $W=5.00, p>0.05$). 
This suggests that \textbf{AR} does not introduce additional cognitive effort for participants to control their virtual motion during interactions.
Participants reported a stronger sense of embodiment ($M=5.13, SD=0.60$) and agency ($M=5.25, SD=0.43$) in \textbf{AR} compared to \textbf{SR}, where embodiment ($M=4.13, SD=0.92$, $W=2.50, p<0.05$) and agency ($M=4.13, SD=0.92$, $W=3.00, p<0.05$) were rated lower. 
This can be attributed to the reduced possibility of detecting the redirection in \textbf{AR}, which enhanced participants' sense of control and immersion. 
In contrast, the frequent detection of redirection in \textbf{SR} reduced their sense of agency and embodiment.

These results suggest that our technique effectively adapts the redirection magnitude to the visual stimuli, aligning with the predicted noticeability from our computational model. 
This demonstrate the potential benefits and capabilities of the model in enhancing redirection interactions.

\section{Discussion}
In this paper, we investigated the effects of visual stimuli on to what extent users notice inconsistencies in their physical movements versus avatar movements. 
We further contribute a regression model that computes the noticeability of redirection under various visual stimuli, based on users' gaze behavioral data.
With the model, we constructed two applications in realistic scenarios with different types of visual stimuli to demonstrate the potential advantages and extensions of our method.
In the following, we discuss possible extensions to our model, as well as limitations and future work.

\subsection{Redirection and visual stimuli}
While prior work~\cite{li2022modeling, feick2023investigating, feick2021visuo} explored how the properties (such as magnitude, direction, location) of redirection influenced its noticeabiltiy, we investigated the noticeability under visual stimuli in this paper.
However, we acknowledge that the redirection properties and the visual stimuli may affect the noticeability in different manners. 
The redirection properties could determine the upper and lower bounds of redirection noticeability, while the visual stimuli can only reduce the noticeability in a limited range.
For subtle redirection that are barely noticeable even when the user is focused on their body movements, adjusting visual stimuli does not significantly alter the noticeability.
Similarly, users will likely notice salient redirection even with a glance, unless the redirection is completely out of their field of view.
Therefore, in this paper, we fixed the redirection magnitude to be 20 degrees (as a control variable), for which the resulting noticeability ranged from approximately 20\% to 80\%.
This relatively large range enables us to quantify the impacts of visual stimuli extensively.
However, we believe that exploring the interaction effect of redirection properties and visual stimuli and combining their influence on noticeability could be important and interesting future work.

\subsection{Diverse visual stimuli}
In this paper, we used several abstract visual stimuli and changed their intensity in our user studies.
We acknowledge that beyond these static visual stimuli tested in this paper, there exist various complicated visual stimuli in realistic use cases.
For instance, a moving object or a wiggling notification may also affect users' gaze behavior and therefore influence the noticeability of redirection.
We consider visual stimuli appearing and staying at a static location to be a standard design paradigm in presenting notifications (e.g., highlighting app icons when new messages are received) on desktop~\cite{muller2023notification}, VR~\cite{rzayev2019notification}, and AR interfaces~\cite{lee2023effects}.
Though we validated our model on new type of abstract visual stimuli and in realistic scenarios,
we acknowledge that verifying the generalizability of our regression model on motion-based or other more complicated visual stimuli is an important future work.
We expect that our research methodology and the presented gaze behavioral patterns can also apply to the investigation of other visual stimuli.

Besides, the visual stimuli investigated in this paper primarily served as external cues for object selection or observation, rather than being directly related to users' body movements. 
In scenarios such as motion training and learning, users may observe their body movements through a mirror or from a third-person perspective, making redirected motion part of the visual stimuli. 
This raises an open question of how to decouple redirection from visual stimuli to investigate their specific influence on the noticeability of redirection. 
We acknowledge this as an important direction for future research.

Furthermore, in more realistic usage scenarios, the stimuli could be in different formats, including instant notifications, environmental events, or even the user's implicit observation of the virtual scene.
We acknowledge that in such cases, 
different behavioral patterns or even physiological signals, such as EEG signals and heart rate variations, can also be indicative of the noticeability of redirection.
We expect that our research method can be adopted to explore further the behavioral patterns that reflect the noticeability in a more realistic setting.


\subsection{User awareness and adaptation}

In our user studies, we hide the true purpose from the participants by disguising it as an accuracy evaluation for a motion tracking system.
The consideration was to mitigate the potential bias of users being aware of the existence of redirection, which might nudge them to be more attuned to or hyper-aware of the redirection.
In addition, our study lasted at most 40 minutes with multiple breaks, which allowed users to regain their perception of their physical movements and prevent fatigue.
However, if a long-term redirection is applied in real-life applications, users might become desensitized to the redirection gradually.
Users may adapt to the redirection after noticing them multiple times and assume that the redirection exists consistently, which could reduce the noticeability of the redirection.
We argue that the regression model should take into account the user's awareness and their ability to adapt their interaction behavior to continue computing the noticeability accurately.




\subsection{Limitations and Future Work}

In our user studies, we treated the ending arm poses that we applied redirection on as a control variable.
We clustered 25 arm poses from the CMU MoCap dataset~\cite{CMUMocap} that are common poses in real-life activities, randomly selecting and testing one of them in each trial.
This enabled us to average the impacts of different arm poses and focus on the influence of visual stimuli on redirection noticeability.
However, we acknowledge that the selected pose set is still limited in size compared to the amount of arm poses that are possible to perform in real life.
We regard extending our study to include more arm poses and apply redirection on other body parts as important future work.

We implemented the regression model with the data from 12 users and evaluated it with another 24 new users.
The results showed that our model could compute the noticeability accurately with new users while they experienced novel visual stimuli that never appeared in the training set.
However, we envision that a personalized model could improve the regression performance by collecting more data from the same user and capturing their unique behavioral patterns more accurately.
In addition, as we primarily focus on modeling the relationship between the visual stimuli and the redirection noticeability, we adopted SVR in the implementation of the regression model as it is relatively stable and did not overfit.
We note that when applying the findings into real life applications, more advanced regression/classification methods (e.g., deep learning models) and more fine-tuned parameters are worth exploring to optimize the regression performance.
As past work has demonstrated the relationship between hand redirection noticeability and users' physiological data~\cite{feick2023investigating}, we will explore how to add physiological data into the regression model in the future.

We investigated the effects of visual stimuli on noticeability and implemented a regression model with highly-controlled study designs and abstract visual stimuli.
Our goal was to study whether and how visual stimuli affects noticeability by controlling the factors and showcasing the potential applications that can benefit from our model.
We regard it as important future work to investigate the effect in a field study with more realistic tasks.
We will also further generalize our contribution with a longitudinal study to consider how users adapt their interaction patterns to redirection over time.




\section{Conclusion}

In this paper, we investigated and modeled the effects of visual stimuli on the noticeability of redirection using users' gaze behaviors in VR.
We first conducted a confirmation study to verify if users' noticeability of redirection was affected by visual stimuli and whether their gaze behaviors were correlated with the noticeability results.
After confirming that visual stimuli could influence the noticeability of redirection, we conducted a data collection study with refined visual stimuli.
With the data, we built up a regression model and selected effective features to compute the noticeability of redirection based on gaze behavior data, achieving an accuracy of 0.011 MSE.
We then evaluated our model on unseen visual stimuli with 24 new users and results suggested that our prediction model could generalize to new visual stimuli.
We then implemented an adaptive redirection technique based on our model and conducted a proof-of-concept study comparing it to static redirection technique.
Results suggested that participants felt less physical demanding while kept a high sense of body ownership using the adaptive redirection technique based on our model.
We believe that our model could support more effective and immersive redirection interactions in VR.

\begin{acks}
We thank Yu Jiang for her research insights.
This work is supported by the National Key Research and Development Program of China under Grant No.2024YFB2808803 and the Natural Science Foundation of China under Grant No. 62102221, 62132010, 62472244, the Tsinghua University Initiative Scientific Research Program, and the Undergraduate Education Innovation Grants, Tsinghua University.
\end{acks}

\bibliographystyle{ACM-Reference-Format}
\bibliography{reference}


\begin{thebibliography}{72}


\ifx \showCODEN    \undefined \def \showCODEN     #1{\unskip}     \fi
\ifx \showDOI      \undefined \def \showDOI       #1{#1}\fi
\ifx \showISBNx    \undefined \def \showISBNx     #1{\unskip}     \fi
\ifx \showISBNxiii \undefined \def \showISBNxiii  #1{\unskip}     \fi
\ifx \showISSN     \undefined \def \showISSN      #1{\unskip}     \fi
\ifx \showLCCN     \undefined \def \showLCCN      #1{\unskip}     \fi
\ifx \shownote     \undefined \def \shownote      #1{#1}          \fi
\ifx \showarticletitle \undefined \def \showarticletitle #1{#1}   \fi
\ifx \showURL      \undefined \def \showURL       {\relax}        \fi
\providecommand\bibfield[2]{#2}
\providecommand\bibinfo[2]{#2}
\providecommand\natexlab[1]{#1}
\providecommand\showeprint[2][]{arXiv:#2}

\bibitem[Abtahi and Follmer(2018)]%
        {abtahi2018visuo}
\bibfield{author}{\bibinfo{person}{Parastoo Abtahi} {and} \bibinfo{person}{Sean Follmer}.} \bibinfo{year}{2018}\natexlab{}.
\newblock \showarticletitle{Visuo-haptic illusions for improving the perceived performance of shape displays}. In \bibinfo{booktitle}{\emph{Proceedings of the 2018 CHI Conference on Human Factors in Computing Systems}}. \bibinfo{pages}{1--13}.
\newblock


\bibitem[Annerer-Walcher et~al\mbox{.}(2021)]%
        {annerer2021reliably}
\bibfield{author}{\bibinfo{person}{Sonja Annerer-Walcher}, \bibinfo{person}{Simon~M Ceh}, \bibinfo{person}{Felix Putze}, \bibinfo{person}{Marvin Kampen}, \bibinfo{person}{Christof K{\"o}rner}, {and} \bibinfo{person}{Mathias Benedek}.} \bibinfo{year}{2021}\natexlab{}.
\newblock \showarticletitle{How reliably do eye parameters indicate internal versus external attentional focus?}
\newblock \bibinfo{journal}{\emph{Cognitive Science}} \bibinfo{volume}{45}, \bibinfo{number}{4} (\bibinfo{year}{2021}), \bibinfo{pages}{e12977}.
\newblock


\bibitem[Azmandian et~al\mbox{.}(2016a)]%
        {azmandian2016haptic}
\bibfield{author}{\bibinfo{person}{Mahdi Azmandian}, \bibinfo{person}{Mark Hancock}, \bibinfo{person}{Hrvoje Benko}, \bibinfo{person}{Eyal Ofek}, {and} \bibinfo{person}{Andrew~D Wilson}.} \bibinfo{year}{2016}\natexlab{a}.
\newblock \showarticletitle{Haptic retargeting: Dynamic repurposing of passive haptics for enhanced virtual reality experiences}. In \bibinfo{booktitle}{\emph{Proceedings of the 2016 chi conference on human factors in computing systems}}. \bibinfo{pages}{1968--1979}.
\newblock


\bibitem[Azmandian et~al\mbox{.}(2016b)]%
        {HapticRetargeting}
\bibfield{author}{\bibinfo{person}{Mahdi Azmandian}, \bibinfo{person}{Mark Hancock}, \bibinfo{person}{Hrvoje Benko}, \bibinfo{person}{Eyal Ofek}, {and} \bibinfo{person}{Andrew~D. Wilson}.} \bibinfo{year}{2016}\natexlab{b}.
\newblock \showarticletitle{Haptic Retargeting: Dynamic Repurposing of Passive Haptics for Enhanced Virtual Reality Experiences}. In \bibinfo{booktitle}{\emph{Proceedings of the 2016 CHI Conference on Human Factors in Computing Systems}} (San Jose, California, USA) \emph{(\bibinfo{series}{CHI '16})}. \bibinfo{publisher}{Association for Computing Machinery}, \bibinfo{address}{New York, NY, USA}, \bibinfo{pages}{1968–1979}.
\newblock
\showISBNx{9781450333627}
\urldef\tempurl%
\url{https://doi.org/10.1145/2858036.2858226}
\showDOI{\tempurl}


\bibitem[Benedek et~al\mbox{.}(2017)]%
        {benedek2017eye}
\bibfield{author}{\bibinfo{person}{Mathias Benedek}, \bibinfo{person}{Robert Stoiser}, \bibinfo{person}{Sonja Walcher}, {and} \bibinfo{person}{Christof K{\"o}rner}.} \bibinfo{year}{2017}\natexlab{}.
\newblock \showarticletitle{Eye behavior associated with internally versus externally directed cognition}.
\newblock \bibinfo{journal}{\emph{Frontiers in psychology}}  \bibinfo{volume}{8} (\bibinfo{year}{2017}), \bibinfo{pages}{1092}.
\newblock


\bibitem[Bixler and D’Mello(2016)]%
        {bixler2016automatic}
\bibfield{author}{\bibinfo{person}{Robert Bixler} {and} \bibinfo{person}{Sidney D’Mello}.} \bibinfo{year}{2016}\natexlab{}.
\newblock \showarticletitle{Automatic gaze-based user-independent detection of mind wandering during computerized reading}.
\newblock \bibinfo{journal}{\emph{User Modeling and User-Adapted Interaction}}  \bibinfo{volume}{26} (\bibinfo{year}{2016}), \bibinfo{pages}{33--68}.
\newblock


\bibitem[Burns et~al\mbox{.}(2006)]%
        {burns2006hand}
\bibfield{author}{\bibinfo{person}{Eric Burns}, \bibinfo{person}{Sharif Razzaque}, \bibinfo{person}{Abigail~T Panter}, \bibinfo{person}{Mary~C Whitton}, \bibinfo{person}{Matthew~R McCallus}, {and} \bibinfo{person}{Frederick~P Brooks~Jr}.} \bibinfo{year}{2006}\natexlab{}.
\newblock \showarticletitle{The hand is more easily fooled than the eye: Users are more sensitive to visual interpenetration than to visual-proprioceptive discrepancy}.
\newblock \bibinfo{journal}{\emph{Presence: teleoperators \& virtual environments}} \bibinfo{volume}{15}, \bibinfo{number}{1} (\bibinfo{year}{2006}), \bibinfo{pages}{1--15}.
\newblock


\bibitem[Chen et~al\mbox{.}(2021)]%
        {chen2021towards}
\bibfield{author}{\bibinfo{person}{Lu Chen}, \bibinfo{person}{Sida Peng}, {and} \bibinfo{person}{Xiaowei Zhou}.} \bibinfo{year}{2021}\natexlab{}.
\newblock \showarticletitle{Towards efficient and photorealistic 3d human reconstruction: a brief survey}.
\newblock \bibinfo{journal}{\emph{Visual Informatics}} \bibinfo{volume}{5}, \bibinfo{number}{4} (\bibinfo{year}{2021}), \bibinfo{pages}{11--19}.
\newblock


\bibitem[Cheng et~al\mbox{.}(2017)]%
        {cheng2017sparse}
\bibfield{author}{\bibinfo{person}{Lung-Pan Cheng}, \bibinfo{person}{Eyal Ofek}, \bibinfo{person}{Christian Holz}, \bibinfo{person}{Hrvoje Benko}, {and} \bibinfo{person}{Andrew~D Wilson}.} \bibinfo{year}{2017}\natexlab{}.
\newblock \showarticletitle{Sparse haptic proxy: Touch feedback in virtual environments using a general passive prop}. In \bibinfo{booktitle}{\emph{Proceedings of the 2017 CHI Conference on Human Factors in Computing Systems}}. \bibinfo{pages}{3718--3728}.
\newblock


\bibitem[Dominjon et~al\mbox{.}(2005)]%
        {dominjon2005influence}
\bibfield{author}{\bibinfo{person}{Lionel Dominjon}, \bibinfo{person}{Anatole L{\'e}cuyer}, \bibinfo{person}{J-M Burkhardt}, \bibinfo{person}{Paul Richard}, {and} \bibinfo{person}{Simon Richir}.} \bibinfo{year}{2005}\natexlab{}.
\newblock \showarticletitle{Influence of control/display ratio on the perception of mass of manipulated objects in virtual environments}. In \bibinfo{booktitle}{\emph{IEEE Proceedings. VR 2005. Virtual Reality, 2005.}} IEEE, \bibinfo{pages}{19--25}.
\newblock


\bibitem[Duchowski et~al\mbox{.}(2018)]%
        {duchowski2018index}
\bibfield{author}{\bibinfo{person}{Andrew~T Duchowski}, \bibinfo{person}{Krzysztof Krejtz}, \bibinfo{person}{Izabela Krejtz}, \bibinfo{person}{Cezary Biele}, \bibinfo{person}{Anna Niedzielska}, \bibinfo{person}{Peter Kiefer}, \bibinfo{person}{Martin Raubal}, {and} \bibinfo{person}{Ioannis Giannopoulos}.} \bibinfo{year}{2018}\natexlab{}.
\newblock \showarticletitle{The index of pupillary activity: Measuring cognitive load vis-{\`a}-vis task difficulty with pupil oscillation}. In \bibinfo{booktitle}{\emph{Proceedings of the 2018 CHI conference on human factors in computing systems}}. \bibinfo{pages}{1--13}.
\newblock


\bibitem[Faber et~al\mbox{.}(2018)]%
        {faber2018automated}
\bibfield{author}{\bibinfo{person}{Myrthe Faber}, \bibinfo{person}{Robert Bixler}, {and} \bibinfo{person}{Sidney~K D’Mello}.} \bibinfo{year}{2018}\natexlab{}.
\newblock \showarticletitle{An automated behavioral measure of mind wandering during computerized reading}.
\newblock \bibinfo{journal}{\emph{Behavior Research Methods}}  \bibinfo{volume}{50} (\bibinfo{year}{2018}), \bibinfo{pages}{134--150}.
\newblock


\bibitem[Feick et~al\mbox{.}(2021)]%
        {feick2021visuo}
\bibfield{author}{\bibinfo{person}{Martin Feick}, \bibinfo{person}{Niko Kleer}, \bibinfo{person}{Andr{\'e} Zenner}, \bibinfo{person}{Anthony Tang}, {and} \bibinfo{person}{Antonio Kr{\"u}ger}.} \bibinfo{year}{2021}\natexlab{}.
\newblock \showarticletitle{Visuo-haptic illusions for linear translation and stretching using physical proxies in virtual reality}. In \bibinfo{booktitle}{\emph{Proceedings of the 2021 CHI Conference on Human Factors in Computing Systems}}. \bibinfo{pages}{1--13}.
\newblock


\bibitem[Feick et~al\mbox{.}(2023)]%
        {feick2023investigating}
\bibfield{author}{\bibinfo{person}{Martin Feick}, \bibinfo{person}{Kora~P Regitz}, \bibinfo{person}{Anthony Tang}, \bibinfo{person}{Tobias Jungbluth}, \bibinfo{person}{Maurice Rekrut}, {and} \bibinfo{person}{Antonio Kr{\"u}ger}.} \bibinfo{year}{2023}\natexlab{}.
\newblock \showarticletitle{Investigating Noticeable Hand Redirection in Virtual Reality using Physiological and Interaction Data}. In \bibinfo{booktitle}{\emph{2023 IEEE Conference Virtual Reality and 3D User Interfaces (VR)}}. IEEE, \bibinfo{pages}{194--204}.
\newblock


\bibitem[Feick et~al\mbox{.}(2024)]%
        {feick2024impact}
\bibfield{author}{\bibinfo{person}{Martin Feick}, \bibinfo{person}{Andr{\'e} Zenner}, \bibinfo{person}{Simon Seibert}, \bibinfo{person}{Anthony Tang}, {and} \bibinfo{person}{Antonio Kr{\"u}ger}.} \bibinfo{year}{2024}\natexlab{}.
\newblock \showarticletitle{The Impact of Avatar Completeness on Embodiment and the Detectability of Hand Redirection in Virtual Reality}. In \bibinfo{booktitle}{\emph{Proceedings of the CHI Conference on Human Factors in Computing Systems}}. \bibinfo{pages}{1--9}.
\newblock


\bibitem[Feuchtner and M{\"u}ller(2018)]%
        {feuchtner2018ownershift}
\bibfield{author}{\bibinfo{person}{Tiare Feuchtner} {and} \bibinfo{person}{J{\"o}rg M{\"u}ller}.} \bibinfo{year}{2018}\natexlab{}.
\newblock \showarticletitle{Ownershift: Facilitating overhead interaction in virtual reality with an ownership-preserving hand space shift}. In \bibinfo{booktitle}{\emph{Proceedings of the 31st Annual ACM Symposium on User Interface Software and Technology}}. \bibinfo{pages}{31--43}.
\newblock


\bibitem[Frees et~al\mbox{.}(2007)]%
        {frees2007prism}
\bibfield{author}{\bibinfo{person}{Scott Frees}, \bibinfo{person}{G~Drew Kessler}, {and} \bibinfo{person}{Edwin Kay}.} \bibinfo{year}{2007}\natexlab{}.
\newblock \showarticletitle{PRISM interaction for enhancing control in immersive virtual environments}.
\newblock \bibinfo{journal}{\emph{ACM Transactions on Computer-Human Interaction (TOCHI)}} \bibinfo{volume}{14}, \bibinfo{number}{1} (\bibinfo{year}{2007}), \bibinfo{pages}{2--es}.
\newblock


\bibitem[Gale(1997)]%
        {gale1997human}
\bibfield{author}{\bibinfo{person}{Alastair~G Gale}.} \bibinfo{year}{1997}\natexlab{}.
\newblock \showarticletitle{Human response to visual stimuli}.
\newblock In \bibinfo{booktitle}{\emph{The perception of visual information}}. \bibinfo{publisher}{Springer}, \bibinfo{pages}{127--147}.
\newblock


\bibitem[Gibson(1933)]%
        {gibson1933adaptation}
\bibfield{author}{\bibinfo{person}{James~J Gibson}.} \bibinfo{year}{1933}\natexlab{}.
\newblock \showarticletitle{Adaptation, after-effect and contrast in the perception of curved lines.}
\newblock \bibinfo{journal}{\emph{Journal of experimental psychology}} \bibinfo{volume}{16}, \bibinfo{number}{1} (\bibinfo{year}{1933}), \bibinfo{pages}{1}.
\newblock


\bibitem[Gillon et~al\mbox{.}(2024)]%
        {gillon2024responses}
\bibfield{author}{\bibinfo{person}{Colleen~J Gillon}, \bibinfo{person}{Jason~E Pina}, \bibinfo{person}{J{\'e}r{\^o}me~A Lecoq}, \bibinfo{person}{Ruweida Ahmed}, \bibinfo{person}{Yazan~N Billeh}, \bibinfo{person}{Shiella Caldejon}, \bibinfo{person}{Peter Groblewski}, \bibinfo{person}{Timothy~M Henley}, \bibinfo{person}{Eric Lee}, \bibinfo{person}{Jennifer Luviano}, {et~al\mbox{.}}} \bibinfo{year}{2024}\natexlab{}.
\newblock \showarticletitle{Responses to pattern-violating visual stimuli evolve differently over days in somata and distal apical dendrites}.
\newblock \bibinfo{journal}{\emph{Journal of Neuroscience}} \bibinfo{volume}{44}, \bibinfo{number}{5} (\bibinfo{year}{2024}).
\newblock


\bibitem[Gonzalez et~al\mbox{.}(2022)]%
        {gonzalez2022model}
\bibfield{author}{\bibinfo{person}{Eric~J Gonzalez}, \bibinfo{person}{Elyse~DZ Chase}, \bibinfo{person}{Pramod Kotipalli}, {and} \bibinfo{person}{Sean Follmer}.} \bibinfo{year}{2022}\natexlab{}.
\newblock \showarticletitle{A Model Predictive Control Approach for Reach Redirection in Virtual Reality}. In \bibinfo{booktitle}{\emph{Proceedings of the 2022 CHI Conference on Human Factors in Computing Systems}}. \bibinfo{pages}{1--15}.
\newblock


\bibitem[Gonzalez and Follmer(2023)]%
        {gonzalez2023sensorimotor}
\bibfield{author}{\bibinfo{person}{Eric~J Gonzalez} {and} \bibinfo{person}{Sean Follmer}.} \bibinfo{year}{2023}\natexlab{}.
\newblock \showarticletitle{Sensorimotor Simulation of Redirected Reaching using Stochastic Optimal Feedback Control}. In \bibinfo{booktitle}{\emph{Proceedings of the 2023 CHI Conference on Human Factors in Computing Systems}}. \bibinfo{pages}{1--17}.
\newblock


\bibitem[Gonzalez-Franco and Lanier(2017)]%
        {gonzalez2017model}
\bibfield{author}{\bibinfo{person}{Mar Gonzalez-Franco} {and} \bibinfo{person}{Jaron Lanier}.} \bibinfo{year}{2017}\natexlab{}.
\newblock \showarticletitle{Model of illusions and virtual reality}.
\newblock \bibinfo{journal}{\emph{Frontiers in psychology}}  \bibinfo{volume}{8} (\bibinfo{year}{2017}), \bibinfo{pages}{1125}.
\newblock


\bibitem[Gonzalez-Franco et~al\mbox{.}(2020)]%
        {gonzalez2020rocketbox}
\bibfield{author}{\bibinfo{person}{Mar Gonzalez-Franco}, \bibinfo{person}{Eyal Ofek}, \bibinfo{person}{Ye Pan}, \bibinfo{person}{Angus Antley}, \bibinfo{person}{Anthony Steed}, \bibinfo{person}{Bernhard Spanlang}, \bibinfo{person}{Antonella Maselli}, \bibinfo{person}{Domna Banakou}, \bibinfo{person}{Nuria Pelechano}, \bibinfo{person}{Sergio Orts-Escolano}, {et~al\mbox{.}}} \bibinfo{year}{2020}\natexlab{}.
\newblock \showarticletitle{The rocketbox library and the utility of freely available rigged avatars}.
\newblock \bibinfo{journal}{\emph{Frontiers in virtual reality}}  \bibinfo{volume}{1} (\bibinfo{year}{2020}), \bibinfo{pages}{20}.
\newblock


\bibitem[Grosvenor and Grosvenor(2007)]%
        {grosvenor2007primary}
\bibfield{author}{\bibinfo{person}{Theodore Grosvenor} {and} \bibinfo{person}{Theodore~P Grosvenor}.} \bibinfo{year}{2007}\natexlab{}.
\newblock \bibinfo{booktitle}{\emph{Primary care optometry}}.
\newblock \bibinfo{publisher}{Elsevier health sciences}.
\newblock


\bibitem[Gutwin et~al\mbox{.}(2017)]%
        {gutwin2017peripheral}
\bibfield{author}{\bibinfo{person}{Carl Gutwin}, \bibinfo{person}{Andy Cockburn}, {and} \bibinfo{person}{Ashley Coveney}.} \bibinfo{year}{2017}\natexlab{}.
\newblock \showarticletitle{Peripheral popout: The influence of visual angle and stimulus intensity on popout effects}. In \bibinfo{booktitle}{\emph{Proceedings of the 2017 CHI conference on human factors in computing systems}}. \bibinfo{pages}{208--219}.
\newblock


\bibitem[Holmqvist et~al\mbox{.}(2011)]%
        {holmqvist2011eye}
\bibfield{author}{\bibinfo{person}{Kenneth Holmqvist}, \bibinfo{person}{Marcus Nystr{\"o}m}, \bibinfo{person}{Richard Andersson}, \bibinfo{person}{Richard Dewhurst}, \bibinfo{person}{Halszka Jarodzka}, {and} \bibinfo{person}{Joost Van~de Weijer}.} \bibinfo{year}{2011}\natexlab{}.
\newblock \bibinfo{booktitle}{\emph{Eye tracking: A comprehensive guide to methods and measures}}.
\newblock \bibinfo{publisher}{oup Oxford}.
\newblock


\bibitem[Hutt et~al\mbox{.}(2017)]%
        {hutt2017gaze}
\bibfield{author}{\bibinfo{person}{Stephen Hutt}, \bibinfo{person}{Jessica Hardey}, \bibinfo{person}{Robert Bixler}, \bibinfo{person}{Angela Stewart}, \bibinfo{person}{Evan Risko}, {and} \bibinfo{person}{Sidney~K D'Mello}.} \bibinfo{year}{2017}\natexlab{}.
\newblock \showarticletitle{Gaze-Based Detection of Mind Wandering during Lecture Viewing.}
\newblock \bibinfo{journal}{\emph{International Educational Data Mining Society}} (\bibinfo{year}{2017}).
\newblock


\bibitem[Hutt et~al\mbox{.}(2016)]%
        {hutt2016eyes}
\bibfield{author}{\bibinfo{person}{Stephen Hutt}, \bibinfo{person}{Caitlin Mills}, \bibinfo{person}{Shelby White}, \bibinfo{person}{Patrick~J Donnelly}, {and} \bibinfo{person}{Sidney~K D'Mello}.} \bibinfo{year}{2016}\natexlab{}.
\newblock \showarticletitle{The Eyes Have It: Gaze-Based Detection of Mind Wandering during Learning with an Intelligent Tutoring System.}
\newblock \bibinfo{journal}{\emph{International Educational Data Mining Society}} (\bibinfo{year}{2016}).
\newblock


\bibitem[Kohli(2010)]%
        {redirectedtouching}
\bibfield{author}{\bibinfo{person}{Luv Kohli}.} \bibinfo{year}{2010}\natexlab{}.
\newblock \showarticletitle{Redirected touching: Warping space to remap passive haptics}. In \bibinfo{booktitle}{\emph{2010 IEEE Symposium on 3D User Interfaces (3DUI)}}. \bibinfo{pages}{129--130}.
\newblock
\urldef\tempurl%
\url{https://doi.org/10.1109/3DUI.2010.5444703}
\showDOI{\tempurl}


\bibitem[Kohli et~al\mbox{.}(2012)]%
        {kohli2012redirected}
\bibfield{author}{\bibinfo{person}{Luv Kohli}, \bibinfo{person}{Mary~C Whitton}, {and} \bibinfo{person}{Frederick~P Brooks}.} \bibinfo{year}{2012}\natexlab{}.
\newblock \showarticletitle{Redirected touching: The effect of warping space on task performance}. In \bibinfo{booktitle}{\emph{2012 IEEE Symposium on 3D User Interfaces (3DUI)}}. IEEE, \bibinfo{pages}{105--112}.
\newblock


\bibitem[Krakowczyk et~al\mbox{.}(2023)]%
        {pymovements}
\bibfield{author}{\bibinfo{person}{Daniel~G. Krakowczyk}, \bibinfo{person}{David~R. Reich}, \bibinfo{person}{Jakob Chwastek}, \bibinfo{person}{Deborah~N. Jakobi}, \bibinfo{person}{Paul Prasse}, \bibinfo{person}{Assunta Süss}, \bibinfo{person}{Oleksii Turuta}, \bibinfo{person}{Paweł Kasprowski}, {and} \bibinfo{person}{Lena~A. Jäger}.} \bibinfo{year}{2023}\natexlab{}.
\newblock \showarticletitle{pymovements: A Python Package for Processing Eye Movement Data}. In \bibinfo{booktitle}{\emph{2023 Symposium on Eye Tracking Research and Applications}} (Tubingen, Germany) \emph{(\bibinfo{series}{ETRA '23})}. \bibinfo{publisher}{Association for Computing Machinery}, \bibinfo{address}{New York, NY, USA}.
\newblock
\showISBNx{979-8-4007-0150-4/23/05}
\urldef\tempurl%
\url{https://doi.org/10.1145/3588015.3590134}
\showDOI{\tempurl}


\bibitem[Krejtz et~al\mbox{.}(2018)]%
        {krejtz2018eye}
\bibfield{author}{\bibinfo{person}{Krzysztof Krejtz}, \bibinfo{person}{Andrew~T Duchowski}, \bibinfo{person}{Anna Niedzielska}, \bibinfo{person}{Cezary Biele}, {and} \bibinfo{person}{Izabela Krejtz}.} \bibinfo{year}{2018}\natexlab{}.
\newblock \showarticletitle{Eye tracking cognitive load using pupil diameter and microsaccades with fixed gaze}.
\newblock \bibinfo{journal}{\emph{PloS one}} \bibinfo{volume}{13}, \bibinfo{number}{9} (\bibinfo{year}{2018}), \bibinfo{pages}{e0203629}.
\newblock


\bibitem[Lab(2021)]%
        {CMUMocap}
\bibfield{author}{\bibinfo{person}{CMU~Graphics Lab}.} \bibinfo{year}{2021}\natexlab{}.
\newblock \bibinfo{title}{CMU Graphics Lab Motion Capture Database}.
\newblock \bibinfo{howpublished}{\url{http://mocap.cs.cmu.edu/}}.
\newblock
\newblock
\shownote{Online; accessed September 2023}.


\bibitem[Langbehn et~al\mbox{.}(2017)]%
        {RDWroom-scale}
\bibfield{author}{\bibinfo{person}{Eike Langbehn}, \bibinfo{person}{Paul Lubos}, \bibinfo{person}{Gerd Bruder}, {and} \bibinfo{person}{Frank Steinicke}.} \bibinfo{year}{2017}\natexlab{}.
\newblock \showarticletitle{Application of redirected walking in room-scale VR}. In \bibinfo{booktitle}{\emph{2017 IEEE Virtual Reality (VR)}}. \bibinfo{pages}{449--450}.
\newblock
\urldef\tempurl%
\url{https://doi.org/10.1109/VR.2017.7892373}
\showDOI{\tempurl}


\bibitem[L{\'e}cuyer(2009)]%
        {lecuyer2009simulating}
\bibfield{author}{\bibinfo{person}{Anatole L{\'e}cuyer}.} \bibinfo{year}{2009}\natexlab{}.
\newblock \showarticletitle{Simulating haptic feedback using vision: A survey of research and applications of pseudo-haptic feedback}.
\newblock \bibinfo{journal}{\emph{Presence: Teleoperators and Virtual Environments}} \bibinfo{volume}{18}, \bibinfo{number}{1} (\bibinfo{year}{2009}), \bibinfo{pages}{39--53}.
\newblock


\bibitem[Lee et~al\mbox{.}(2023)]%
        {lee2023effects}
\bibfield{author}{\bibinfo{person}{Hyunjin Lee}, \bibinfo{person}{Sunyoung Bang}, {and} \bibinfo{person}{Woontack Woo}.} \bibinfo{year}{2023}\natexlab{}.
\newblock \showarticletitle{Effects of coordinate system and position of AR notification while walking}.
\newblock \bibinfo{journal}{\emph{Virtual Reality}} \bibinfo{volume}{27}, \bibinfo{number}{2} (\bibinfo{year}{2023}), \bibinfo{pages}{829--848}.
\newblock


\bibitem[Lee et~al\mbox{.}(2015)]%
        {lee2015enlarging}
\bibfield{author}{\bibinfo{person}{Yongseok Lee}, \bibinfo{person}{Inyoung Jang}, {and} \bibinfo{person}{Dongjun Lee}.} \bibinfo{year}{2015}\natexlab{}.
\newblock \showarticletitle{Enlarging just noticeable differences of visual-proprioceptive conflict in VR using haptic feedback}. In \bibinfo{booktitle}{\emph{2015 IEEE World Haptics Conference (WHC)}}. IEEE, \bibinfo{pages}{19--24}.
\newblock


\bibitem[Leek(2001)]%
        {leek2001adaptive}
\bibfield{author}{\bibinfo{person}{Marjorie~R Leek}.} \bibinfo{year}{2001}\natexlab{}.
\newblock \showarticletitle{Adaptive procedures in psychophysical research}.
\newblock \bibinfo{journal}{\emph{Perception \& psychophysics}} \bibinfo{volume}{63}, \bibinfo{number}{8} (\bibinfo{year}{2001}), \bibinfo{pages}{1279--1292}.
\newblock


\bibitem[Li et~al\mbox{.}(2024)]%
        {li2024predicting}
\bibfield{author}{\bibinfo{person}{Zhipeng Li}, \bibinfo{person}{Yi~Fei Cheng}, \bibinfo{person}{Yukang Yan}, {and} \bibinfo{person}{David Lindlbauer}.} \bibinfo{year}{2024}\natexlab{}.
\newblock \showarticletitle{Predicting the Noticeability of Dynamic Virtual Elements in Virtual Reality}. In \bibinfo{booktitle}{\emph{Proceedings of the CHI Conference on Human Factors in Computing Systems}}. \bibinfo{pages}{1--17}.
\newblock


\bibitem[Li et~al\mbox{.}(2022)]%
        {li2022modeling}
\bibfield{author}{\bibinfo{person}{Zhipeng Li}, \bibinfo{person}{Yu Jiang}, \bibinfo{person}{Yihao Zhu}, \bibinfo{person}{Ruijia Chen}, \bibinfo{person}{Ruolin Wang}, \bibinfo{person}{Yuntao Wang}, \bibinfo{person}{Yukang Yan}, {and} \bibinfo{person}{Yuanchun Shi}.} \bibinfo{year}{2022}\natexlab{}.
\newblock \showarticletitle{Modeling the Noticeability of User-Avatar Movement Inconsistency for Sense of Body Ownership Intervention}.
\newblock \bibinfo{journal}{\emph{Proceedings of the ACM on Interactive, Mobile, Wearable and Ubiquitous Technologies}} \bibinfo{volume}{6}, \bibinfo{number}{2} (\bibinfo{year}{2022}), \bibinfo{pages}{1--26}.
\newblock


\bibitem[Lindlbauer et~al\mbox{.}(2019)]%
        {lindlbauer2019context}
\bibfield{author}{\bibinfo{person}{David Lindlbauer}, \bibinfo{person}{Anna~Maria Feit}, {and} \bibinfo{person}{Otmar Hilliges}.} \bibinfo{year}{2019}\natexlab{}.
\newblock \showarticletitle{Context-aware online adaptation of mixed reality interfaces}. In \bibinfo{booktitle}{\emph{Proceedings of the 32nd annual ACM symposium on user interface software and technology}}. \bibinfo{pages}{147--160}.
\newblock


\bibitem[Lohse et~al\mbox{.}(2019)]%
        {lohse2019leveraging}
\bibfield{author}{\bibinfo{person}{Andreas~L Lohse}, \bibinfo{person}{Christoffer~K Kj{\ae}r}, \bibinfo{person}{Ervin Hamulic}, \bibinfo{person}{Ingrid~GA Lima}, \bibinfo{person}{Tilde~H Jensen}, \bibinfo{person}{Luis~E Bruni}, {and} \bibinfo{person}{Niels~C Nilsson}.} \bibinfo{year}{2019}\natexlab{}.
\newblock \showarticletitle{Leveraging change blindness for haptic remapping in virtual environments}. In \bibinfo{booktitle}{\emph{2019 IEEE 5th Workshop on Everyday Virtual Reality (WEVR)}}. IEEE, \bibinfo{pages}{1--5}.
\newblock


\bibitem[Marwecki et~al\mbox{.}(2019)]%
        {marwecki2019mise}
\bibfield{author}{\bibinfo{person}{Sebastian Marwecki}, \bibinfo{person}{Andrew~D Wilson}, \bibinfo{person}{Eyal Ofek}, \bibinfo{person}{Mar Gonzalez~Franco}, {and} \bibinfo{person}{Christian Holz}.} \bibinfo{year}{2019}\natexlab{}.
\newblock \showarticletitle{Mise-unseen: Using eye tracking to hide virtual reality scene changes in plain sight}. In \bibinfo{booktitle}{\emph{Proceedings of the 32nd Annual ACM Symposium on User Interface Software and Technology}}. \bibinfo{pages}{777--789}.
\newblock


\bibitem[Maslych et~al\mbox{.}(2023)]%
        {maslych2023effective}
\bibfield{author}{\bibinfo{person}{Mykola Maslych}, \bibinfo{person}{Eugene~Matthew Taranta}, \bibinfo{person}{Mostafa Aldilati}, {and} \bibinfo{person}{Joseph~J Laviola}.} \bibinfo{year}{2023}\natexlab{}.
\newblock \showarticletitle{Effective 2D Stroke-based Gesture Augmentation for RNNs}. In \bibinfo{booktitle}{\emph{Proceedings of the 2023 CHI Conference on Human Factors in Computing Systems}}. \bibinfo{pages}{1--13}.
\newblock


\bibitem[McInnes et~al\mbox{.}(2017)]%
        {leland2017hdbscan}
\bibfield{author}{\bibinfo{person}{Leland McInnes}, \bibinfo{person}{John Healy}, {and} \bibinfo{person}{Steve Astels}.} \bibinfo{year}{2017}\natexlab{}.
\newblock \showarticletitle{hdbscan: Hierarchical density based clustering}.
\newblock \bibinfo{journal}{\emph{J. Open Source Softw.}} \bibinfo{volume}{2}, \bibinfo{number}{11} (\bibinfo{year}{2017}), \bibinfo{pages}{205}.
\newblock
\urldef\tempurl%
\url{https://doi.org/10.21105/joss.00205}
\showDOI{\tempurl}


\bibitem[Mills et~al\mbox{.}(2016)]%
        {mills2016automatic}
\bibfield{author}{\bibinfo{person}{Caitlin Mills}, \bibinfo{person}{Robert Bixler}, \bibinfo{person}{Xinyi Wang}, {and} \bibinfo{person}{Sidney~K D'Mello}.} \bibinfo{year}{2016}\natexlab{}.
\newblock \showarticletitle{Automatic Gaze-Based Detection of Mind Wandering during Narrative Film Comprehension.}
\newblock \bibinfo{journal}{\emph{International Educational Data Mining Society}} (\bibinfo{year}{2016}).
\newblock


\bibitem[Montano~Murillo et~al\mbox{.}(2017)]%
        {montano2017erg}
\bibfield{author}{\bibinfo{person}{Roberto~A Montano~Murillo}, \bibinfo{person}{Sriram Subramanian}, {and} \bibinfo{person}{Diego Martinez~Plasencia}.} \bibinfo{year}{2017}\natexlab{}.
\newblock \showarticletitle{Erg-O: Ergonomic optimization of immersive virtual environments}. In \bibinfo{booktitle}{\emph{Proceedings of the 30th annual ACM symposium on user interface software and technology}}. \bibinfo{pages}{759--771}.
\newblock


\bibitem[M\"{u}ller et~al\mbox{.}(2022)]%
        {muller2023notification}
\bibfield{author}{\bibinfo{person}{Philipp M\"{u}ller}, \bibinfo{person}{Sander Staal}, \bibinfo{person}{Mihai B\^{a}ce}, {and} \bibinfo{person}{Andreas Bulling}.} \bibinfo{year}{2022}\natexlab{}.
\newblock \showarticletitle{Designing for Noticeability: Understanding the Impact of Visual Importance on Desktop Notifications}. In \bibinfo{booktitle}{\emph{Proceedings of the 2022 CHI Conference on Human Factors in Computing Systems}} (New Orleans, LA, USA) \emph{(\bibinfo{series}{CHI '22})}. \bibinfo{publisher}{Association for Computing Machinery}, \bibinfo{address}{New York, NY, USA}, Article \bibinfo{articleno}{472}, \bibinfo{numpages}{13}~pages.
\newblock
\showISBNx{9781450391573}
\urldef\tempurl%
\url{https://doi.org/10.1145/3491102.3501954}
\showDOI{\tempurl}


\bibitem[Ogawa et~al\mbox{.}(2020)]%
        {ogawa2020effect}
\bibfield{author}{\bibinfo{person}{Nami Ogawa}, \bibinfo{person}{Takuji Narumi}, {and} \bibinfo{person}{Michitaka Hirose}.} \bibinfo{year}{2020}\natexlab{}.
\newblock \showarticletitle{Effect of avatar appearance on detection thresholds for remapped hand movements}.
\newblock \bibinfo{journal}{\emph{IEEE transactions on visualization and computer graphics}} \bibinfo{volume}{27}, \bibinfo{number}{7} (\bibinfo{year}{2020}), \bibinfo{pages}{3182--3197}.
\newblock


\bibitem[Patras et~al\mbox{.}(2022)]%
        {patras2022body}
\bibfield{author}{\bibinfo{person}{Cristian Patras}, \bibinfo{person}{Mantas Cibulskis}, {and} \bibinfo{person}{Niels~Christian Nilsson}.} \bibinfo{year}{2022}\natexlab{}.
\newblock \showarticletitle{Body warping versus change blindness remapping: A comparison of two approaches to repurposing haptic proxies for virtual reality}. In \bibinfo{booktitle}{\emph{2022 IEEE Conference on Virtual Reality and 3D User Interfaces (VR)}}. IEEE, \bibinfo{pages}{205--212}.
\newblock


\bibitem[Peck and Gonzalez-Franco(2021)]%
        {peck2021avatar}
\bibfield{author}{\bibinfo{person}{Tabitha~C Peck} {and} \bibinfo{person}{Mar Gonzalez-Franco}.} \bibinfo{year}{2021}\natexlab{}.
\newblock \showarticletitle{Avatar embodiment. a standardized questionnaire}.
\newblock \bibinfo{journal}{\emph{Frontiers in Virtual Reality}}  \bibinfo{volume}{1} (\bibinfo{year}{2021}), \bibinfo{pages}{575943}.
\newblock


\bibitem[Perkhofer and Lehner(2019)]%
        {perkhofer2019using}
\bibfield{author}{\bibinfo{person}{Lisa Perkhofer} {and} \bibinfo{person}{Othmar Lehner}.} \bibinfo{year}{2019}\natexlab{}.
\newblock \showarticletitle{Using gaze behavior to measure cognitive load}. In \bibinfo{booktitle}{\emph{Information Systems and Neuroscience: NeuroIS Retreat 2018}}. Springer, \bibinfo{pages}{73--83}.
\newblock


\bibitem[Poupyrev et~al\mbox{.}(1996)]%
        {poupyrev1996go}
\bibfield{author}{\bibinfo{person}{Ivan Poupyrev}, \bibinfo{person}{Mark Billinghurst}, \bibinfo{person}{Suzanne Weghorst}, {and} \bibinfo{person}{Tadao Ichikawa}.} \bibinfo{year}{1996}\natexlab{}.
\newblock \showarticletitle{The go-go interaction technique: non-linear mapping for direct manipulation in VR}. In \bibinfo{booktitle}{\emph{Proceedings of the 9th annual ACM symposium on User interface software and technology}}. \bibinfo{pages}{79--80}.
\newblock


\bibitem[Razzaque(2005)]%
        {razzaque2005redirected}
\bibfield{author}{\bibinfo{person}{Sharif Razzaque}.} \bibinfo{year}{2005}\natexlab{}.
\newblock \bibinfo{booktitle}{\emph{Redirected walking}}.
\newblock \bibinfo{publisher}{The University of North Carolina at Chapel Hill}.
\newblock


\bibitem[Rietzler et~al\mbox{.}(2020)]%
        {rietzler2020telewalk}
\bibfield{author}{\bibinfo{person}{Michael Rietzler}, \bibinfo{person}{Martin Deubzer}, \bibinfo{person}{Thomas Dreja}, {and} \bibinfo{person}{Enrico Rukzio}.} \bibinfo{year}{2020}\natexlab{}.
\newblock \showarticletitle{Telewalk: Towards free and endless walking in room-scale virtual reality}. In \bibinfo{booktitle}{\emph{Proceedings of the 2020 CHI Conference on Human Factors in Computing Systems}}. \bibinfo{pages}{1--9}.
\newblock


\bibitem[Rock and Victor(1964)]%
        {rock1964vision}
\bibfield{author}{\bibinfo{person}{Irvin Rock} {and} \bibinfo{person}{Jack Victor}.} \bibinfo{year}{1964}\natexlab{}.
\newblock \showarticletitle{Vision and touch: An experimentally created conflict between the two senses}.
\newblock \bibinfo{journal}{\emph{Science}} \bibinfo{volume}{143}, \bibinfo{number}{3606} (\bibinfo{year}{1964}), \bibinfo{pages}{594--596}.
\newblock


\bibitem[Rzayev et~al\mbox{.}(2019)]%
        {rzayev2019notification}
\bibfield{author}{\bibinfo{person}{Rufat Rzayev}, \bibinfo{person}{Sven Mayer}, \bibinfo{person}{Christian Krauter}, {and} \bibinfo{person}{Niels Henze}.} \bibinfo{year}{2019}\natexlab{}.
\newblock \showarticletitle{Notification in VR: The Effect of Notification Placement, Task and Environment}. In \bibinfo{booktitle}{\emph{Proceedings of the Annual Symposium on Computer-Human Interaction in Play}} (Barcelona, Spain) \emph{(\bibinfo{series}{CHI PLAY '19})}. \bibinfo{publisher}{Association for Computing Machinery}, \bibinfo{address}{New York, NY, USA}, \bibinfo{pages}{199–211}.
\newblock
\showISBNx{9781450366885}
\urldef\tempurl%
\url{https://doi.org/10.1145/3311350.3347190}
\showDOI{\tempurl}


\bibitem[Samad et~al\mbox{.}(2019)]%
        {samad2019pseudo}
\bibfield{author}{\bibinfo{person}{Majed Samad}, \bibinfo{person}{Elia Gatti}, \bibinfo{person}{Anne Hermes}, \bibinfo{person}{Hrvoje Benko}, {and} \bibinfo{person}{Cesare Parise}.} \bibinfo{year}{2019}\natexlab{}.
\newblock \showarticletitle{Pseudo-haptic weight: Changing the perceived weight of virtual objects by manipulating control-display ratio}. In \bibinfo{booktitle}{\emph{Proceedings of the 2019 CHI Conference on Human Factors in Computing Systems}}. \bibinfo{pages}{1--13}.
\newblock


\bibitem[Shakhnarovich(2005)]%
        {shakhnarovich05learning}
\bibfield{author}{\bibinfo{person}{Gregory Shakhnarovich}.} \bibinfo{year}{2005}\natexlab{}.
\newblock \emph{\bibinfo{title}{Learning task-specific similarity}}.
\newblock \bibinfo{thesistype}{Ph.\,D. Dissertation}. \bibinfo{school}{Massachusetts Institute of Technology, Cambridge, MA, {USA}}.
\newblock
\urldef\tempurl%
\url{http://hdl.handle.net/1721.1/36138}
\showURL{%
\tempurl}


\bibitem[Steinicke et~al\mbox{.}(2009)]%
        {steinicke2009estimation}
\bibfield{author}{\bibinfo{person}{Frank Steinicke}, \bibinfo{person}{Gerd Bruder}, \bibinfo{person}{Jason Jerald}, \bibinfo{person}{Harald Frenz}, {and} \bibinfo{person}{Markus Lappe}.} \bibinfo{year}{2009}\natexlab{}.
\newblock \showarticletitle{Estimation of detection thresholds for redirected walking techniques}.
\newblock \bibinfo{journal}{\emph{IEEE transactions on visualization and computer graphics}} \bibinfo{volume}{16}, \bibinfo{number}{1} (\bibinfo{year}{2009}), \bibinfo{pages}{17--27}.
\newblock


\bibitem[Suma et~al\mbox{.}(2010)]%
        {suma2010exploiting}
\bibfield{author}{\bibinfo{person}{Evan~A Suma}, \bibinfo{person}{Seth Clark}, \bibinfo{person}{Samantha~L Finkelstein}, {and} \bibinfo{person}{Zachary Wartell}.} \bibinfo{year}{2010}\natexlab{}.
\newblock \showarticletitle{Exploiting change blindness to expand walkable space in a virtual environment}. In \bibinfo{booktitle}{\emph{2010 IEEE Virtual Reality Conference (VR)}}. IEEE, \bibinfo{pages}{305--306}.
\newblock


\bibitem[Suma et~al\mbox{.}(2012)]%
        {suma2012impossible}
\bibfield{author}{\bibinfo{person}{Evan~A Suma}, \bibinfo{person}{Zachary Lipps}, \bibinfo{person}{Samantha Finkelstein}, \bibinfo{person}{David~M Krum}, {and} \bibinfo{person}{Mark Bolas}.} \bibinfo{year}{2012}\natexlab{}.
\newblock \showarticletitle{Impossible spaces: Maximizing natural walking in virtual environments with self-overlapping architecture}.
\newblock \bibinfo{journal}{\emph{IEEE Transactions on Visualization and Computer Graphics}} \bibinfo{volume}{18}, \bibinfo{number}{4} (\bibinfo{year}{2012}), \bibinfo{pages}{555--564}.
\newblock


\bibitem[Wang et~al\mbox{.}(2019)]%
        {wang2019exploring}
\bibfield{author}{\bibinfo{person}{Jiahui Wang}, \bibinfo{person}{Pavlo Antonenko}, \bibinfo{person}{Mehmet Celepkolu}, \bibinfo{person}{Yerika Jimenez}, \bibinfo{person}{Ethan Fieldman}, {and} \bibinfo{person}{Ashley Fieldman}.} \bibinfo{year}{2019}\natexlab{}.
\newblock \showarticletitle{Exploring relationships between eye tracking and traditional usability testing data}.
\newblock \bibinfo{journal}{\emph{International Journal of Human--Computer Interaction}} \bibinfo{volume}{35}, \bibinfo{number}{6} (\bibinfo{year}{2019}), \bibinfo{pages}{483--494}.
\newblock


\bibitem[Wentzel et~al\mbox{.}(2020)]%
        {wentzel2020improving}
\bibfield{author}{\bibinfo{person}{Johann Wentzel}, \bibinfo{person}{Greg d'Eon}, {and} \bibinfo{person}{Daniel Vogel}.} \bibinfo{year}{2020}\natexlab{}.
\newblock \showarticletitle{Improving virtual reality ergonomics through reach-bounded non-linear input amplification}. In \bibinfo{booktitle}{\emph{Proceedings of the 2020 CHI Conference on Human Factors in Computing Systems}}. \bibinfo{pages}{1--12}.
\newblock


\bibitem[Yu and Bowman(2020)]%
        {yu2020pseudo}
\bibfield{author}{\bibinfo{person}{Run Yu} {and} \bibinfo{person}{Doug~A Bowman}.} \bibinfo{year}{2020}\natexlab{}.
\newblock \showarticletitle{Pseudo-haptic display of mass and mass distribution during object rotation in virtual reality}.
\newblock \bibinfo{journal}{\emph{IEEE transactions on visualization and computer graphics}} \bibinfo{volume}{26}, \bibinfo{number}{5} (\bibinfo{year}{2020}), \bibinfo{pages}{2094--2103}.
\newblock


\bibitem[Zagermann et~al\mbox{.}(2016)]%
        {zagermann2016measuring}
\bibfield{author}{\bibinfo{person}{Johannes Zagermann}, \bibinfo{person}{Ulrike Pfeil}, {and} \bibinfo{person}{Harald Reiterer}.} \bibinfo{year}{2016}\natexlab{}.
\newblock \showarticletitle{Measuring cognitive load using eye tracking technology in visual computing}. In \bibinfo{booktitle}{\emph{Proceedings of the sixth workshop on beyond time and errors on novel evaluation methods for visualization}}. \bibinfo{pages}{78--85}.
\newblock


\bibitem[Zenner et~al\mbox{.}(2023)]%
        {zenner2023detectability}
\bibfield{author}{\bibinfo{person}{Andr{\'e} Zenner}, \bibinfo{person}{Chiara Karr}, \bibinfo{person}{Martin Feick}, \bibinfo{person}{Oscar Ariza}, {and} \bibinfo{person}{Antonio Kr{\"u}ger}.} \bibinfo{year}{2023}\natexlab{}.
\newblock \showarticletitle{The Detectability of Saccadic Hand Offset in Virtual Reality}. In \bibinfo{booktitle}{\emph{Proceedings of the 29th ACM Symposium on Virtual Reality Software and Technology}}. \bibinfo{pages}{1--2}.
\newblock


\bibitem[Zenner et~al\mbox{.}(2024)]%
        {zenner2024beyond}
\bibfield{author}{\bibinfo{person}{Andr{\'e} Zenner}, \bibinfo{person}{Chiara Karr}, \bibinfo{person}{Martin Feick}, \bibinfo{person}{Oscar Ariza}, {and} \bibinfo{person}{Antonio Kr{\"u}ger}.} \bibinfo{year}{2024}\natexlab{}.
\newblock \showarticletitle{Beyond the Blink: Investigating Combined Saccadic \& Blink-Suppressed Hand Redirection in Virtual Reality}. In \bibinfo{booktitle}{\emph{Proceedings of the CHI Conference on Human Factors in Computing Systems}}. \bibinfo{pages}{1--14}.
\newblock


\bibitem[Zenner and Kr{\"u}ger(2019)]%
        {zenner2019estimating}
\bibfield{author}{\bibinfo{person}{Andr{\'e} Zenner} {and} \bibinfo{person}{Antonio Kr{\"u}ger}.} \bibinfo{year}{2019}\natexlab{}.
\newblock \showarticletitle{Estimating detection thresholds for desktop-scale hand redirection in virtual reality}. In \bibinfo{booktitle}{\emph{2019 IEEE Conference on Virtual Reality and 3D User Interfaces (VR)}}. IEEE, \bibinfo{pages}{47--55}.
\newblock


\bibitem[Zenner et~al\mbox{.}(2021)]%
        {zenner2021blink}
\bibfield{author}{\bibinfo{person}{Andr{\'e} Zenner}, \bibinfo{person}{Kora~Persephone Regitz}, {and} \bibinfo{person}{Antonio Kr{\"u}ger}.} \bibinfo{year}{2021}\natexlab{}.
\newblock \showarticletitle{Blink-suppressed hand redirection}. In \bibinfo{booktitle}{\emph{2021 IEEE Virtual Reality and 3D User Interfaces (VR)}}. IEEE, \bibinfo{pages}{75--84}.
\newblock


\bibitem[Zhao and Follmer(2018)]%
        {zhao2018functional}
\bibfield{author}{\bibinfo{person}{Yiwei Zhao} {and} \bibinfo{person}{Sean Follmer}.} \bibinfo{year}{2018}\natexlab{}.
\newblock \showarticletitle{A functional optimization based approach for continuous 3d retargeted touch of arbitrary, complex boundaries in haptic virtual reality}. In \bibinfo{booktitle}{\emph{Proceedings of the 2018 CHI Conference on Human Factors in Computing Systems}}. \bibinfo{pages}{1--12}.
\newblock


\end{thebibliography}

\appendix

\clearpage
\newpage

\section{Target pose samples}
\label{appendix:poses}

In this section, we present a set of 25 target poses sampled from the CMU MoCap dataset~\cite{CMUMocap}, using the HDBSCAN clustering algorithm~\cite{leland2017hdbscan} for pose selection. 
The samples represent a diverse range of human body postures.

\begin{figure}[h]
    \centering
    \includegraphics[width=0.95\linewidth]{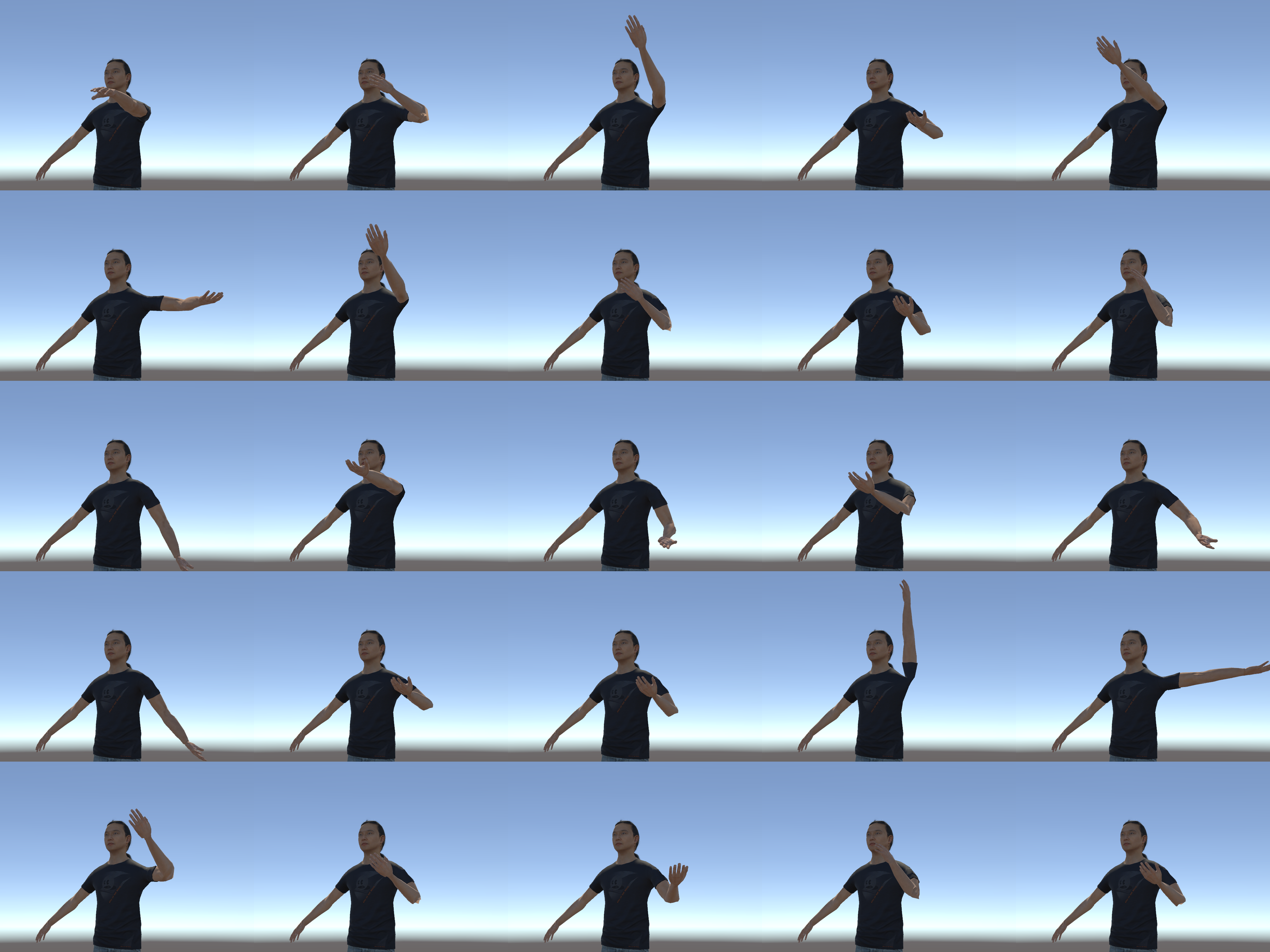}
    \caption{Sampled target poses from the CMU MoCap dataset, clustered using HDBSCAN.}
    \label{figure:targetpose}
\end{figure}

\end{document}